\documentclass[fleqn,usenatbib]{mnras}

\usepackage{newtxtext,newtxmath}

\usepackage[T1]{fontenc}
\usepackage{ae,aecompl}

\usepackage{graphicx}	
\usepackage{amsmath}	
\usepackage{ulem}

\usepackage{soul}	
\usepackage{xcolor}

\newcommand{\beq}{\begin{equation}}
\newcommand{\eeq}{\end{equation}}
\newcommand{\bea}{\begin{eqnarray}}
\newcommand{\eea}{\end{eqnarray}}
\newcommand{\nn}{\nonumber}

\usepackage{color}
\usepackage{subfigure}
\usepackage{multirow}
\usepackage{float}
\usepackage{nameref}
\usepackage{natbib}
\usepackage{soul}	
\usepackage{xcolor}

\title[Particle acceleration in winds of star clusters]{Particle acceleration in winds of star clusters}

\author[]{
	G. Morlino $^{1}$\thanks{E-mail: giovanni.morlino@inaf.it},
	P. Blasi$^{2,3}$\thanks{E-mail: pasquale.blasi@gssi.it},
	E. Peretti$^{4,2}$\thanks{E-mail: peretti@nbi.ku.dk} \&
	P. Cristofari$^{2,3}$\thanks{E-mail: pierre.cristofari@gssi.it}		
	\\
	$^{1}$ INAF/Osservatorio Astrofisico di Arcetri, Largo E. Fermi, 5 - 50125 Firenze, Italy\\
	$^{2}$ Gran Sasso Science Institute (INFN), Viale F. Crispi 7 - 67100 L' Aquila, Italy\\
	$^{3}$ INFN-Laboratori Nazionali del Gran Sasso, Via G. Acitelli 22, Assergi (AQ), Italy \\
	$^{4}$ Niels Bohr International Academy, Niels Bohr Institute, University of Copenhagen, Blegdamsvej 17, DK-2100 Copenhagen, Denmark \\
}

\date{Accepted XXX. Received YYY; in original form ZZZ}

\pubyear{2021}

\begin{document}

\label{firstpage}
\pagerange{\pageref{firstpage}--\pageref{lastpage}}
\maketitle

\begin{abstract}
The origin of cosmic rays in our Galaxy remains a subject of active debate. While supernova remnant shocks are often invoked as the sites of acceleration, it is now widely accepted that the difficulties of such sources in reaching PeV energies are daunting and it seems likely that only a subclass of rare remnants can satisfy the necessary conditions. Moreover the spectra of cosmic rays escaping the remnants have a complex shape that is not obviously the same as the spectra observed at the Earth. Here we investigate the process of particle acceleration at the termination shock that develops in the bubble excavated by star clusters' winds in the interstellar medium. While the main limitation to the maximum energy in supernova remnants comes from the need for effective wave excitation upstream so as to confine particles in the near-shock region and speed up the acceleration process, at the termination shock of star clusters the confinement of particles upstream in guaranteed by the geometry of the problem. We develop a theory of diffusive shock acceleration at such shock and we find that the maximum energy may reach the PeV region for powerful clusters in the high end of the luminosity tail for these sources. A crucial role in this problem is played by the dissipation of energy in the wind to magnetic perturbations. Under reasonable conditions the spectrum of the accelerated particles has a power law shape with a slope $4\div 4.3$, in agreement with what is required based upon standard models of cosmic ray transport in the Galaxy.  
\end{abstract}




\section{Introduction}
\label{sec:introduction}

The standard scenario for the acceleration of cosmic rays (CRs) in the Galaxy is based upon the so-called supernova remnant (SNR) paradigm: acceleration occurs through the mechanism of diffusive acceleration at the shock fronts produced as a result of the supersonic motion of the supernova ejecta in the surrounding medium (DSA) \cite[]{blandford}. Soon after the proposal of DSA as the chief mechanism for particle energization in SNRs, it became clear that the maximum energy of accelerated particles is exceedingly low unless waves are effectively excited upstream of the shock, due to the same particles that are being accelerated \cite[]{Lagage1,Lagage2}. Even in the presence of this mechanism of self-confinement, based on the excitation of a resonant streaming instability \cite[]{Kulsrud:1969p1936}, the maximum energy in typical SNRs can hardly exceed $\sim 100$ TeV, more than one order of magnitude below the energy of the knee. Recently two new pieces were added to the puzzle: on the observational side, it was found that virtually all young SNRs observed in the X-ray band are characterised by the presence of bright non-thermal thin X-ray rims, coincident with the position of the forward shock \cite[see][for reviews]{Ballet2006,Vink:2012p2755}. The morphology of the rims allowed us to estimate the strength of the magnetic field in the shock region, thereby showing that such field is about $\sim 100$ times larger than the typical fields in the interstellar medium (ISM). From the theoretical point of view, \cite{bell2004} discovered the existence of a non-resonant quasi-purely growing mode excited by CRs upstream of the shock, able to account for the strong field observed in the X-rays. It was soon realised that the growth of the instability would saturate when the magnetic field energy density and the energy density in the form of escaping particles become comparable \cite[]{Schure:2013p3169}. This condition would in principle lead to much larger values of the maximum energy of accelerated particles than with resonant streaming instability alone. However, since the initial excitement for this discovery, it has become clear that when applied to conditions specific of supernova remnants of type Ia and core collapse, although the instantaneous maximum energy may exceed the energy of the knee, after integration over the history of the SNR, the spectrum of CRs released into the ISM would show an effective maximum energy typically in the $10-100$ TeV range \cite[]{pierre}. 
The only possible exception to this conclusion applies to powerful ($\gtrsim 5 \times 10^{51}$ erg), rare ($\sim 1/10^{4}$ years) core collapse SNRs, with relatively small ejecta mass (few solar masses), for which the maximum energy can indeed reach PeV energies. The overall spectrum of CRs released in the ISM by each of the classes of SN explosions mentioned above seems bumpy and unlike the relatively smooth spectrum observed at the Earth. Although these problems and difficulties might only suggest that our theoretical approaches to the origin of CRs in SNRs are too simplistic, they have also stimulated the search for alternative sources of CRs, with special care for those that produce a spectrum extending to the knee energy. In this context, stellar clusters~\citep{reimer2006}, OB associations~\citep{bykov2001,volk1982}, and supperbubbles~\citep{bykov2001_2,parizot2004} have for instance been proposed.

It has especially been speculated that the winds of massive stars may be a suitable location for the acceleration of CRs \cite[]{Cesarsky-Montmerle:1983,webb1985,Gupta+2018,Bykov+2020}. Moreover, recently the gamma ray emission from the region around a few compact star clusters has been measured, including Westerlund 1 \citep{Abramowski_Wd1:2012}, Westerlund 2 \citep{Yang+2018}, Cygnus cocoon \citep{Ackermann:2011p3159,Aharonian+2019NatAs}, NGC 3603 \citep{Saha_NGC3603:2020}, BDS2003 \citep{HAWC:2021}, W40 \citep{Sun_W40:2020} and 30 Doradus in the LMC \citep{HESS-30Dor:2015}. These observations have been used to infer the spatial distribution of CRs and their energy budget, supporting the scenario in which a sizable fraction of the wind kinetic energy is converted to non thermal particles and, at the same time, maximum energies $> 100$~TeV are reached. These findings would, than, suggest that stellar clusters can substantially contribute to the flux of Galactic CRs.

Further support to such a conclusion comes from the analysis of the $^{22}$Ne/$^{20}$Ne abundance in CRs, which is a factor $\sim 5$ larger than for the solar wind \citep{Binns+:2006}. This result is not easy to accommodate in the framework of particle acceleration at SNR shocks alone \citep{Prantzos2012} while can be more easily accounted for if CRs are at least partly accelerated out of material contained in the winds of massive stars \citep{Gupta+2020}. 

Here we show that the termination shock formed as a result of the interaction of the intense collective wind of the star cluster with the ISM is a potentially interesting site for particle acceleration up to $\sim$PeV energies, for several reasons: first, particle escape from the upstream region (in the direction of the star cluster itself) is forbidden because of the geometry of the problem; 2) if a relatively small fraction ($\sim 10\%$) of the wind kinetic energy is dissipated to magnetic energy, particle diffusion around the shock can be reduced, thereby shortening the acceleration time; 3) if the kinetic luminosity of the star cluster is large enough ($\gtrsim 3\times 10^{38}$ erg/s) then the maximum energy is indeed in the $\sim$PeV range; 4) in rather common situations around the termination shock, the spectrum of accelerated particles may be somewhat steeper than $E^{-2}$, as required by observations of CRs on Galactic scale \cite[]{Evoli1,Evoli2}. 

The article is organised as follows: in \S \ref{sec:bubble} we briefly describe the structure of the environment around the star cluster and the properties of the termination shock where particle acceleration is expected to take place. In \S~\ref{sec:diff} we discuss the diffusion properties of particles inside the wind bubble while in \S~\ref{sec:theory} we describe in detail the solution of the DSA problem at the termination shock and we derive an expression for the maximum energy of accelerated particles. In \S \ref{sec:conc} we summarise our findings and we comment on the possibility that star clusters may in fact be prominent contributors to the flux of CRs in the Galaxy. 
 
\section{The bubble's structure}
\label{sec:bubble}

\begin{figure}
\centering
\includegraphics[width=.35\textwidth]{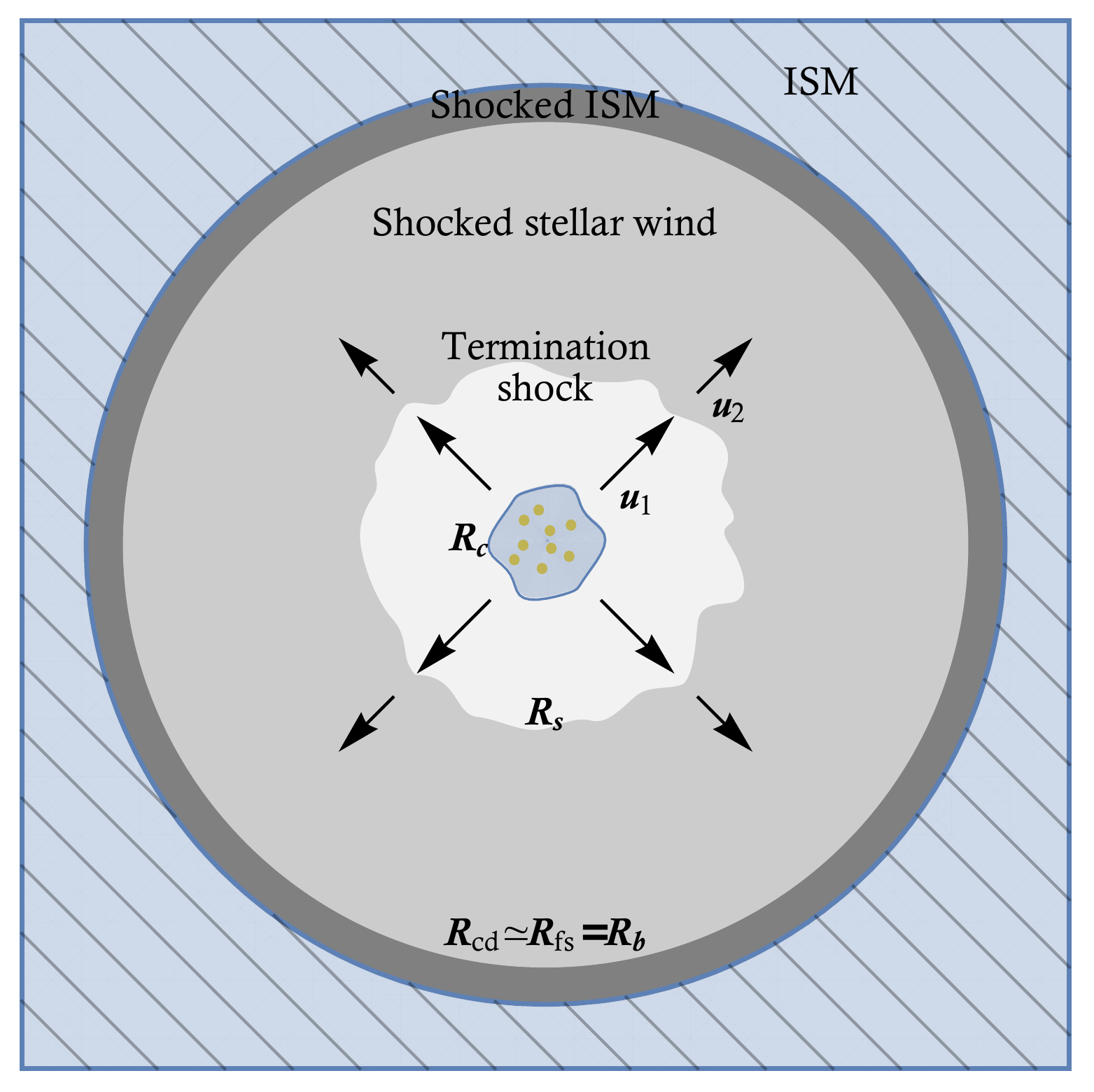}
\caption{Schematic structure of a wind bubble excavated by a star cluster into the ISM: $R_{s}$ marks the position of the termination shock, $R_{\rm cd}$ the contact discontinuity, and $R_{\rm fs}$ the forward shock.}
\label{fig:geometry}
\end{figure}

The bubble excavated by the collective stellar wind launched by the star cluster is schematically illustrated in Fig.~\ref{fig:geometry}: the central part is filled with the wind itself, expanding with a velocity $v_{w}$ and density 
\beq
\rho(r) = \frac{\dot M}{4\pi r^{2} v_{w}},~~~ r > R_{c},
\eeq
where $R_{c}$ is the radius of the core where the stars are concentrated, and $\dot M$ is the rate of mass loss due to the collective wind. The impact of the supersonic wind with the ISM, assumed here to have a constant density $\rho_{0}$, produces a forward shock at position $R_{\rm fs}$, while the shocked wind is bound by a termination shock, at a location $R_{s}$. The shocked ISM and the shocked wind are separated by a contact discontinuity at $R_{\rm cd}$. The typical cooling timescale of the shocked ISM is only $\sim 10^4$ yr, while the cooling time for the shocked wind is several $10^7$ yr which is comparable with the typical age of these systems \citep{Koo-McKee:1992a, Koo-McKee:1992b}. As a consequence, the wind-blown bubble spends the largest part of its life in a quasi-adiabatic phase, meaning that the shocked wind is adiabatic while the shocked ISM is cold and dense and compressed in a very thin layer, such that we can approximate $R_{\rm cd}\simeq R_{\rm fs} \equiv R_b$. Hence most of the volume of the bubble is filled with the wind and the shocked wind. Below, following \cite{Weaver+1977} and \cite{Gupta+2018} we provide a simple approximation for the position in time of the forward shock (FS) and the termination shock (TS). 
The mass accumulated at the FS while moving in the ISM is $M(R)=\int_{0}^{R}  4\pi r^{2} \rho_{0} dr$, where $\rho_0$ is the external density. The momentum of the material accumulated in the thin shell between $R_{\rm cd}$ and $R_{\rm fs}$ is $M(R)\dot R$ and changes because of the work done by the pressure $P$ in the hot bubble:
\beq
\frac{d}{dt}\left[ M(R)\dot R \right] = 4\pi R^{2} P.
\label{eq:pressure}
\eeq
On the other hand, the energy density in the bubble is $\epsilon = \frac{4}{3}\pi R^{3}\frac{P}{\gamma_{g}-1}$, where $\gamma_{g}$ is the adiabatic index, and it changes according to
\beq
\frac{d}{dt}\left[  \frac{4}{3}\pi R^{3}\frac{P}{\gamma_{g}-1} \right] = L_{w} + 4\pi R^{2} \dot R P - L_{\rm cool},
\label{eq:energy}
\eeq
where $L_{w}=\frac{1}{2}\dot M v_{w}^{2}$ is the wind luminosity and $L_{\rm cool}$ is the cooling rate. In the following, for the purpose of a simple estimate we will neglect this cooling term which is only important at very late times \citep{Koo-McKee:1992a, Stevens-Hartwell:2003}. In general we can assume that $L_{\rm cool}\sim \zeta L_{w}$, so that the results that will be found below will rescale with $L_{w}\to (1-\zeta)L_{w}$. If we look for solutions in the form $R(t)=At^{\alpha}$, it is easy to show, using Eq.~\eqref{eq:pressure}, that $P=\frac{1}{3}A^{2}\alpha \rho_{0}(4\alpha-1)t^{2\alpha-2}$. Replacing this expression in Eq.~\eqref{eq:energy} leads to 
\beq
\frac{4\pi}{9}\frac{A^{5} \alpha \rho_{0} (4\alpha-1)(5\alpha-2)t^{5\alpha-3}}{\gamma_{g}-1} = L_{w} - \frac{4\pi}{3} A^{5} \alpha^{2} \rho_{0} (4\alpha-1)t^{5\alpha-3},
\eeq
which readily implies that $\alpha=3/5$ and 
\beq
A =  \left[ \frac{3\times 5^{3}(\gamma_{g}-1)}{4\pi(63\gamma_{g}-28)} \, \frac{L_{w}}{\rho_{0}} \right]^{1/5} 
\simeq 0.76 \left[\frac{L_{w}}{\rho_{0}} \right]^{1/5}.
\eeq
In the last equality we have used $\gamma_{g}=5/3$. Normalizing to typical values of the parameters, we obtain:
\beq \label{eq:Rbubble}
 R_b(t) = 174  ~ \rho_{1}^{-1/5} L_{37}^{1/5} t_{10}^{3/5} ~\rm pc,
\eeq
where $\rho_{1}$ is the ISM density in the region around the star cluster in units of 1 proton per cm$^{3}$, $L_{37}=L_{w}/(10^{37}\rm erg~s^{-1})$ and $t_{10}$ is the dynamical time in units of 10 million years. It is worth noticing that the shell moves outwards with a velocity $\dot R_b$ that is only a few tens of km/s, thereby being at most transonic. To first approximation, the position of the termination shock can be easily derived by imposing balance between the pressure $P$ and the ram pressure of the wind: 
\beq
 \frac{\dot M v_{w}}{4\pi R_{s}^{2}} = \frac{7}{25}A^{2} \rho_{0} t^{-4/5},
\eeq
which leads to
\beq
 R_{s}= 62 ~ \dot{M}_{-4}^{1/2} v_{8}^{1/2} \rho_{1}^{-3/10} L_{37}^{-1/5} t_{10}^{2/5} ~\rm pc, 
\eeq
where $v_{8}=v_{w}/(1000~\rm km\, s^{-1})$ and $\dot M_{-4}=\dot M/(10^{-4}\rm M_{\odot}\,yr^{-1})$. If we use the definition of $L_{w}$ and we neglect cooling, this expression can be rewritten as:
\beq \label{eq:RTS}
 R_{s}= 48.6 ~ \dot{M}_{-4}^{3/10} v_{8}^{1/10} \rho_{1}^{-3/10} t_{10}^{2/5} ~\rm pc \,.
\eeq
A more accurate calculation \citep{Weaver+1977} shows  that the result above is accurate within $\lesssim 10\%$. We stress again that the speed of the TS in the laboratory frame is very low, so that the entire bubble structure evolves slowly and can be considered as stationary to first approximation. It is worth stressing that the formation of a collective wind occurs only for compact clusters that have a typical cluster size $R_c \ll R_{s}$ \cite[see, e.g.][]{Gupta+2020}.

Typically the core of a massive stellar cluster can contain up to $\sim 100-1000$ stars whose winds interact strongly leading to partial dissipation of kinetic energy of the winds, which may result in generation of turbulent magnetic field in the free expanding wind. This implies that the collective wind outside the core is not expected to have a coherent, spiral-like structure. In the next section we discuss the properties of particle diffusion in such an environment.


\section{Diffusion around the termination shock}
\label{sec:diff}

While for the wind of an individual star it is conceivable to think that the magnetic field retains memory of its spiral structure, this assumption would be untenable for the wind of a star cluster because of the large number of winds that collide and interact in the core region. On the other hand, the fact that winds of individual stars are characterized by different mass loss rates and different velocities results in the interaction among different components which, to some extent, should result in dissipation of the kinetic energy of these winds to thermal and magnetic energy. 
Whether this process of partial equipartition occurs only at the base of the wind region or everywhere in the wind is not clear. Hence, below we will consider a situation in which some fraction $\eta_B$ of the kinetic energy of the wind is transformed to magnetic energy at any radius (\S~\ref{sec:MHD_turbulence}), and we discuss the role of self-generated magnetic fields, due to CR induced instabilities, when the fraction $\eta_B$ is small (\S~\ref{sec:self-generation}). 

\subsection{MHD turbulence}
\label{sec:MHD_turbulence}
Let us assume that a fraction $\eta_B$ of the wind kinetic energy is converted to magnetic turbulence at any location, in such a way that the strength of the turbulent magnetic field can be written as
\beq
  B(r) \approx \frac{1}{r}\left( \frac{1}{2}\eta_{B} \dot M v_{w} \right)^{1/2}.
\eeq
At the location of the termination shock, the strength of the magnetic field reads
\beq \label{eq:B_TS}
B(R_{s})=3.7\times 10^{-6} \eta_{B}^{1/2} \dot M_{-4}^{1/5}v_{8}^{2/5} \rho_{1}^{3/10} t_{10}^{-2/5}~ G.
\eeq
This dissipation of kinetic energy into magnetic energy likely results in turbulence with a typical scale $L_c$ that is expected to be of order the size of the star cluster, $L_c\sim R_c\sim 1\div 2$ pc. If the turbulence evolves following a Kolmogorov cascade, the diffusion coefficient immediately upstream of the termination shock can be estimated as
\begin{flalign}  \label{eq:DKol}
 D(E)\approx\frac{1}{3} r_{L}(p) v \left(\frac{r_{L}(p)}{L_{c}}\right)^{-2/3} = 
 2\times 10^{26} 
 \left( \frac{L_c}{1\rm pc}\right)^{2/3} \hspace{0.9cm} \nn \\
  \eta_{B}^{-1/6} \dot M_{-4}^{-1/15} v_{8}^{-2/15} \rho_{1}^{-1/10} t_{10}^{2/15} E_{\rm GeV}^{1/3}~\rm cm^2~s^{-1},
\end{flalign}
where $r_{L}(p)=pc/e B(r)$ is the Larmor radius of particles of momentum $p$ in the magnetic field $B(r)$. The diffusion coefficient decreases inward as $(r/R_{s})^{1/3}$. One can see that the dependence of the diffusion coefficient upon the efficiency of conversion of kinetic energy to magnetic energy, $\eta_{B}$, is very weak. Downstream of the termination shock, it is assumed that the magnetic field is only compressed by the standard factor $\sqrt{11}$, typical of a strong shock, so that $D_{2}\approx 0.67 D_{1}$. Clearly the downstream diffusion coefficient can be smaller than this estimate suggests, if other processes (such as the Richtmyer-Meshkov instability \cite[]{Giacalone:2007p962}) lead to enhanced turbulence behind the shock. 

An order of magnitude for the maximum energy that can be achieved through DSA at the termination shock of the wind can be easily obtained by requiring that the diffusion length of the particles at the highest energy be equal to the radius of the termination shock, $D_{1}(E_{\max}) / v_{w}\approx R_{\rm s}$. This value should be taken with much caution, in that the actual maximum energy can be somewhat smaller depending on the diffusion coefficient downstream of the shock. We will discuss these effects in the next section, where we develop a formal theory of DSA at the termination shock, taking into account the geometry of the problem and the escape of accelerated particles from the bubble. 

The simple criterion discussed above, using Kolmogorov turbulence, leads to:
\beq
 E_{\max} \approx 10^{14} \, \eta_{B}^{1/2} \dot M_{-4}^{11/10} v_{8}^{37/10}\rho_{1}^{-3/5}t_{10}^{4/5}
 \left( \frac{L_c}{2\rm pc}\right)^{-2}~ \rm eV.
\label{eq:maxE}
\eeq

The expression for the diffusion coefficient in Eq.~\eqref{eq:DKol} is valid as long as the Larmor radius of particles is smaller than $L_c$. Using Eq.\eqref{eq:B_TS} this constraint can also be written as:
\beq
 E\lesssim 6.8\times 10^{15} \, \eta_{B}^{1/2} \dot M_{-4}^{1/5} v_{8}^{2/5}\rho_{1}^{3/10}t_{10}^{-2/5}
 \left( \frac{L_c}{2\rm pc}\right)~ \rm eV.
\label{eq:maxELc}
\eeq
For larger energies, $D(E)\propto E^2$, independent of the type of turbulent cascading \cite[see, for instance,][]{Dundovic2020}, and acceleration quickly becomes inefficient. 

Imposing that $E_{\max}$ does not exceed the bound in Eq. \eqref{eq:maxELc} one obtains the additional constraint:
\beq
\dot M_{-4}^{9/10} v_{8}^{33/10}\rho_{1}^{-9/10}t_{10}^{6/5}
\left( \frac{L_c}{1\rm pc}\right)^{-3}\lesssim 69
\label{eq:constraint}	
\eeq
One can see from Eq.~\eqref{eq:maxE} that in order to reach PeV energies, for the reference values of the parameters one needs wind speeds of $\sim 2500$ km/s using $\eta_{B}\sim 0.1$. The constraint in Eq.~\eqref{eq:constraint} implies that the wind speed be $\lesssim 3600$ km/s for the same reference values of the other parameters (notice however the strong dependence upon $L_c$). It follows that a typical star cluster may produce particles with energy in the PeV energy region, but not much larger than that. The dependence of this conclusion upon the spectrum of the turbulence in the wind region is relatively weak: if the turbulence follows a Kraichnan cascading process, such that $D(E) = v/3\, (r_L L_c)^{1/2}$, it can be easily seen that the maximum energy imposed by the condition $D_{1}(E_{\max}) / u_{1}\approx R_{\rm s}$ reads
\beq
 E_{\max}\approx 4\times 10^{14} \, \eta_{B}^{1/2} \dot M_{-4}^{4/5} v_{8}^{13/5}\rho_{1}^{-3/10}t_{10}^{2/5}
 \left( \frac{L_c}{2\rm pc}\right)^{-1}~ \rm eV.
\label{eq:maxEkra}
\eeq
In this case, in order to reach PeV energies one needs wind speeds larger than $\sim 2000$ km s$^{-1}$ for $\eta_{B}\sim 0.1$ and the other parameters chosen at their reference values.

In both cases it appears that massive star clusters characterized by large wind speeds can account for CR acceleration in the knee region, provided turbulence can be developed down to small enough scales to ensure resonant scattering. The time required for such a cascade process to take place can be estimated (at the termination shock) as
\beq
\tau_c\simeq \frac{L_c}{v_A}=2.9~ v_8^{-1} \eta_B^{-1/2} 
\left( \frac{L_c}{2\rm pc}\right) ~\rm kyr,
\eeq
where $v_A= B_0/\sqrt{4\pi \rho} =\eta_B^{1/2}\sqrt{2} \, v_w$ is the Alfv\'en speed (spatially constant in the cold wind). The time $\tau_c$ is clearly much shorter than the dynamical time scale of a star cluster, but it is also required to be shorter than the advection time of the wind across the region between the star cluster and the termination shock, i.e. $t_w = R_{s}/v_w$. For our standard parameters' values we have $\tau_c/t_w = L_c v_w/(R_s v_A) \simeq 0.1$.

For both models of turbulent cascading the dependence of the maximum energy upon wind speed is rather strong ($\sim v_8^{3.7}$ for Kolmogorov and $\sim v_8^{2.6}$ for Kraichnan). This strong dependence is the reason why the maximum energy is in the PeV region only for very fast winds, while rapidly dropping to lower values for slower, most common star cluster winds.

A comment about the expected spectrum of accelerated particles is in order. While DSA at a strong shock almost invariably leads to a spectrum  $f(p) \propto p^{-4}$, independent of the geometry of the shock, multi-wavelengths observations of young SNRs (like Tycho or Cas A) require a proton spectrum $\propto p^{-4.3}$ \cite[]{Caprioli:2011p2134}. Interestingly, the same spectral index is also inferred based upon gamma-ray spectra measured from massive stellar clusters \citep{Aharonian+2019NatAs}, and inferred from CR transport in the Galaxy \cite[]{Evoli1,Evoli2}. From the theoretical point of view, some deviations from the standard predicted spectra are expected when the Alfv\'enic Mach number is finite and of order a few \cite[]{Bell:1978p1344}. In the case discussed above, the magnetic field at the shock is as given in Eq.~\eqref{eq:B_TS} and the Alfv\'en speed can be easily calculated to be $v_{A,1}=v_w\eta_B^{1/2}\sqrt{2}$, for a strong shock. This means that the Alfv\'enic Mach number is $\sim 4.5$. Because of the development of turbulence in the upstream plasma, one can expect that the effective Alfv\'en speed, accounting for the waves moving in all directions, is vanishingly small. On the other hand, as shown by \cite{Caprioli+2020} using hybrid simulations, for self-generated perturbations, downstream of the shock there seems to be a net velocity of these waves in the direction away from the shock. In a parametric form, we can write the mean velocity of the waves downstream as $\bar{v}_{A,2}=\chi \frac{\sqrt{11}}{2\sqrt{2}}\eta_B^{1/2}v_w$, where $\chi=0$ for waves that are symmetrically moving in all directions. 

On a very general ground, the slope of accelerated particles is determined by the effective compression ratio which accounts for the average speed of the scattering centers:
\beq
  \sigma_c = \frac{v_{w,1}}{v_{w,2} + \bar{v}_{A,2}}=\frac{4}{1+4.68\chi \,\eta_B^{1/2}} \,.
\eeq
A spectral slope of 4.3 would require $\sigma_c=3.3$, which in turn would imply $\chi \, \eta_B^{1/2}=4.5\times 10^{-2}$. Using as a reference value $\eta_B \approx 0.1$, this condition translates to $\chi \approx 14\%$. Hence an asymmetry at the level of $\sim 10\div 15\%$ in the modes would be sufficient to produce spectra of accelerated particles somewhat steeper than $p^{-4}$.

\subsection{Self-generated turbulence}
\label{sec:self-generation}

On top of MHD turbulence, some level of magnetic field self-generation is also expected due to the excitation of streaming instability by accelerated particles in the proximity of the termination shock. Below we briefly discuss the resonant and the non-resonant branch of this instability. If the spectrum of accelerated particles is $\sim p^{-4}$, then the resonant instability produces a flat turbulence power spectrum \cite[]{Amato-Blasi2006}:
\beq
  \mathcal{F}_{\rm res} = \left(\frac{\delta B}{B_1} \right)^2
  = \frac{\pi}{2} \frac{\xi_{\rm CR}}{\Lambda} \frac{v_w}{v_A}
  = \frac{\pi}{2} \frac{\xi_{\rm CR}}{\Lambda} (2 \eta_B)^{-1/2},
\eeq
where we introduced $\Lambda=\ln(p_{\max}/m_p c)\sim 13$. Notice that here we are assuming that the self-generated turbulence is produced on top of a large scale field (yet turbulent on smaller scales). This is a rather risky procedure for a few reasons: first, the instability is calculated assuming that there is a regular, well defined field that defines the unperturbed particle trajectories, not a turbulent field. Second, in the presence of pre-existing turbulence, the growth of the instability is quenched, as discussed by \cite{Farmer}. In conclusion, the power spectrum reported above should be considered as an absolute upper limit to the strength of the phenomenon. In any case, one can see that $\mathcal{F}_{\rm res}$ becomes of order unity only for $\eta_B\lesssim 10^{-4}$, a rather small value. In any case the turbulent quenching would make this phenomenon of little impact. 


Contrary to resonant modes, the non-resonant streaming instability \cite[]{bell2004} is allowed to grow only if the energy density in the CR current times $v_w/c$ is smaller than the energy density in the pre-existing magnetic field. In terms of $\eta_B$, the condition for the growth of this instability translates to requiring that 
\beq
\eta_B\lesssim \frac{6\xi_{\rm CR}}{\Lambda}
\frac{v_w}{c},
\eeq
where, as above, we assumed that the spectrum of accelerated particles is $\propto p^{-4}$. For an efficiency of particle acceleration $\xi_{\rm CR}\sim 0.1$, the non resonant modes are excited provided that less than $\sim 10^{-4}$ of the ram pressure of the wind is converted to magnetic turbulence at any given radius in the wind, especially at the location of the termination shock. This implies that even when the instability is allowed to grow, it cannot lead to magnetic fields in excess of those obtained above for $\eta_B\sim 10^{-4}$. Such constraint would limit the maximum energy of accelerated particles to exceedingly low values. Hence, even the growth of the non resonant instability leads to a less optimistic result that the one presented in the case in which a fraction $\eta_B>10^{-4}$ of the ram pressure is converted to magnetic turbulence.

\section{Theory of DSA at the wind termination shock}
\label{sec:theory}

\subsection{General solution} \label{sec:general}
The peculiar geometry of the wind region requires a detailed calculation of the spectrum and spatial distribution of the particles accelerated at the termination shock. Given the quasi-stationary evolution of the wind region, the CR transport is modelled using the standard time independent transport equation in spherical symmetry: 
\bea
  \frac{\partial}{\partial r} \left[ r^2 D(r,p) \frac{\partial f}{\partial r} \right]
  - r^2 u(r) \frac{\partial f}{\partial r} 
  + \frac{d\left[ r^2 u(r) \right]}{d r} \frac{p}{3} \frac{\partial f}{\partial p}  \nn \\
  + \, r^2 Q(z,p) = 0  ,
  \label{eq:transport}
\eea
where $u(r)$ is the plasma speed and $D(r,p)$ is the diffusion coefficient. 

Particle acceleration takes place only at the termination shock, located at $r=R_s$, where particles are injected according to:
\beq
  Q(r,p) = Q_0(p) \delta(r-R_{s}) 
            = \frac{\eta_{\rm inj }n_{1} u_1}{4 \pi p_{\rm inj}^2} \delta(p-p_{\rm inj}) \delta(r-R_{s}),
  \label{eq:Qinj}
\eeq
where $n_{1}$ is the density of the cold wind immediately upstream of the termination shock and $\eta_{\rm inj}$ is the fraction of particle flux that takes part in the acceleration process. 


The solution of the transport equation is found by first solving the equation upstream (unshocked wind) and downstream (shocked wind) and then joining the two at the shock surface, where the solution is $f_s(p)\equiv f(r=R_s,p)$. Because of the spatial dependence of the plasma velocity and the spherical symmetry, the solution is found adopting an iterative procedure, similar to that introduced to treat non-linear DSA \cite[]{2002APh....16..429B,2004APh....21...45B,2005MNRAS.364L..76A,Amato-Blasi2006}.


Let us first consider the upstream region, $r < R_s$.  Integrating Eq.(\ref{eq:transport}) between $0$ and  $r$ and using the boundary condition that there is no net flux at $r=0$, namely $\left[r^2 D\partial_r f - r^2 u f \right]_{r=0}=0$, we get 
\begin{equation}
  \frac{\partial f}{\partial r} = \frac{u(r)}{D(r,p)} f + \frac{ G_1(r,p) }{r^2 D(r,p)}, 
  \label{eq:partial_f1}
\end{equation}
where we introduced the function 
\beq    \label{eq:G1}
  G_1(r,p) = \int_0^{r} f(r',p) \frac{\tilde{q}(r',p)}{3} \frac{d (r^2 u) }{d r'} dr' \,,
\eeq
and we used the identity
\beq  \label{eq:qtilde}
  f + \frac{p}{3} \frac{\partial f}{\partial p} = \frac{f}{3} \frac{\partial \ln(p^3 f)}{\partial \ln p} = - f \frac{\tilde{q}}{3}
  \hspace{1cm}
\eeq
with $\tilde{q} \equiv -\frac{\partial \ln(p^3 f)}{\partial \ln p}$. Eq.(\ref{eq:partial_f1}) can be solved in an implicit way and leads to

\beq \label{eq:sol_f1}
  f_1(r,p) = f_s(p) \, \exp { \left\{-\int_r^{R_s} \frac{u}{D} \left[ 1+ \frac{G_1(r',p)}{r'^2 u f(r',p)} \right]dr' \right\} } \,.
\eeq
The second term in square brackets, containing the function $G_{1}$, is in an implicit form and accounts for adiabatic losses/gains. One can recover the standard solution for the case of plane parallel shocks by imposing $G_1=0$.

The solution in the downstream region, $R_{s}< r < R_b$, where $R_{b}$ is the radius of the bubble, can also be easily found by integrating the transport equation between $R_{s}$ and $r<R_{b}$:
\beq
  \left[ r^2 D \frac{\partial f}{\partial r} \right]_{R_s^+}^{r} 
  - r^2 u f + R_s^2 u_2 f_s  +  G_2(r,p) - G_2(R_s,p) = 0 ,
 \label{eq:tran_int2}  
\eeq
where $u_2 \equiv u(R_s^+)$ and we have defined 
\beq    \label{eq:G2}
  G_2(r,p) = \int_{r}^{R_b} f(r,p) \frac{\tilde{q}}{3} \frac{d (r'^2 u) }{d r'} dr' \,.
\eeq
Using the definition of the escape flux, $ \phi_{\rm esc} \equiv - \left[ r^2 D\partial_r f - r^2 u(r) f\right]_{r=R_{b}}$, we  can write the derivative of $f$ immediately downstream of the shock as
\beq
  \left[ D \frac{\partial f}{\partial r}\right]_{R_s^+} = u_2 f_s 
  	- \frac{\phi_{\rm esc}}{R_s^2} - \frac{G_2(R_s^+,p)}{R_s^2}.
\label{eq:Ddfdr2} 
\eeq
Using Eq.(\ref{eq:Ddfdr2}) into Eq.(\ref{eq:tran_int2}) we get the equation for $f$, which reads
\beq
  \frac{\partial f}{\partial r} = \frac{u}{D} f - \frac{G_2(r,p)}{r^2 D}  - \frac{\phi_{\rm esc}}{r^2 D}  \,.
  \label{eq:partial_f2}
\eeq
The solution of this equation can be written as:
\beq
  f_2(r,p) = \int_{r}^{R_b} dr' \frac{G_2 + \phi_{\rm esc}}{r'^2 D} \exp\left[ - \int_{r}^{r'} \frac{u}{D} dr'' \right] \,,
\eeq
or, if to write it in terms of the distribution function at the shock, $f_s$:
\bea \label{eq:sol_f2}
  f_2(r,p) = \left[ f_s(p) - \int_{R_s}^{r} \frac{G_2 + \phi_{\rm esc}}{r'^2 D} 
  		\exp\left\{- \int_{R_s}^{r'} \frac{u}{D} dr'' \right\} \right]  	\nn \\
 	     \times \exp{ \left\{\int_{R_s}^{r} \frac{u}{D} dr'\right\} }.
\eea
In general, the escape flux $\phi_{\rm esc}$ can be related to $f_s$ solving the transport equation outside of the bubble, as we will show in the next section, although a good approximation to the solution can be obtained assuming free escape from the edge of the bubble.

At this point the two solutions, upstream and downstream can be joined at the shock. Integrating Eq.(\ref{eq:transport}) between $R_{s}^-$ and  $R_{s}^+$ and recalling that the plasma velocity is discontinuous, namely $u(r) \simeq u_2 + (u_1-u_2) \theta(R_s-r)$, the derivative at $r=R_s$ is $d(r^2 u)/dr = - R_s^2 (u_1-u_2) \delta(r-R_s)$, hence one has: 
\beq
  \left[ D \frac{\partial f}{\partial r} \right]_{R_s^+} - 
  \left[ D \frac{\partial f}{\partial r} \right]_{R_s^-}
  - \left( u_1 - u_2\right) \frac{p}{3} \frac{\partial f_s}{\partial p} + Q_0(p)  = 0
\eeq
The first two terms in square brackets can be obtained from Eqs.(\ref{eq:Ddfdr2}) and (\ref{eq:partial_f1}), respectively, giving:
\beq
  \left( u_1 -u_2 \right) \frac{p}{3} \frac{\partial f_s}{\partial p} = 
  - \left( u_1 -u_2 \right) f_s - \frac{ \phi_{\rm esc} + G}{R_s^2}  
  + Q_0
  \label{eq:fs_diff}
\eeq
where $G(p) \equiv G_1(R_{s}, p) + G_2(R_{s},p)$. Eq.(\ref{eq:fs_diff}) is a first order differential equation in $p$ and can be solved in an implicit form once $\phi_{\rm esc}$ is written as a function of $f_s$ using Eq.\eqref{eq:sol_f2}.

\subsection{The case of adiabatic bubble}

In this section we specialise the solution found above in the case of a wind-bubble system to the semi-adiabatic phase, as described in \S~\ref{sec:general}. We assume, as it is usually the case, a constant speed for the cold wind, $u(r < R_s)=u_1=v_{w}$, and a profile  $u(r > R_s)=u_2 (R_{s}/r)^2$ inside the shocked wind region, where $u_2 = u_1/\sigma$ is the velocity immediately downstream of the termination shock and $\sigma$ is the compression ratio. Such a velocity profile implies that the function $G_2=0$, meaning that there are no adiabatic losses in the shocked wind region. Outside the bubble the plasma is assumed to be at rest, $u(r > R_b) = 0$.
Following the discussion in \S~\ref{sec:general}, we  assume that the diffusion coefficient downstream, $D_2(p)$, is spatially constant (because the shocked wind is subsonic) while upstream we allow $D_1$ to have a spatial dependence.

Now, we start simplifying the solution in the downstream from Eq.\eqref{eq:sol_f2}. We first define the function
\beq
  \alpha_2(r,p) \equiv \int_{R_s}^{r} \frac{u(r)}{D_2(p)} dr' 
  = \frac{u_2 R_{s}}{D_2(p)} \left( 1 - \frac{R_s}{r}\right) \,.
\eeq
At the boundary of the bubble we can assume $f_b\ll f_s$ (that will be justified {\it a posteriori}), such that the escaping flux can be obtained from Eq.\eqref{eq:sol_f2} and reads:
\beq  \label{eq:phi_esc}
  \phi_{\rm esc}(p) = R_s^2 \, \frac{u_2 f_s(p)}{1- e^{-\alpha_2(R_b)}} \,,
\eeq
which, inserted back into Eq.~\eqref{eq:sol_f2}, returns the solution in the shocked wind region:
\beq  \label{eq:sol_f2_3}
  f_2(r,p)= f_s(p) \, \frac{1 - e^{\alpha_2(r) - \alpha_2(R_b)} }{1 - e^{-\alpha_2(R_b)}}  \,.
\eeq
The value of $f_b$ can be estimated by solving the transport equation outside the bubble.
Under the assumption that the diffusion coefficient in the ISM, $D_{0}$, is constant,  Eq.\eqref{eq:transport} reduces to $r^2 D_0 \partial_r f = \rm const$. Integrating this equation with the  two boundary conditions $f(R_b,p) = f_b(p)$ and $f(r \rightarrow \infty,p) = 0$, we get 
\beq
  f(r>R_b,p) = f_b(p) R_b/r \,.
\label{eq:f_ISM}
\eeq
The escaping flux evaluated at $R_s^+$ is, then, $ \phi_{\rm esc} \equiv - r^2 D\partial_r f |_{r=R_{b}} = f_b D_0 R_b$. By equating this expression to Eq.~\eqref{eq:phi_esc} we obtain
\beq  \label{eq:f_b_approx}
  f_{b}(p) = \frac{R_s^2}{R_b^2} \frac{ R_b}{D_0} \frac{u_2 f_s(p)}{1- e^{-\alpha_2(R_b)}}.
\eeq
For small momenta, such that $\alpha_2 \ll 1$, the assumption $f_b \ll f_s$ is verified when $D_2 \ll D_0 R_s/R_b$. Assuming that $D_0$ is of the order of the average galactic diffusion coefficient, i.e. $D_{\rm gal} \simeq 3 \times 10^{28} (E/{\rm GeV})^{1/3} \rm cm^2~ s^{-1}$ and comparing such a value with $D_2$ from Eq.~\eqref{eq:DKol}, one can see that the above condition in easily fulfilled. For larger momenta ($\alpha_2 \gtrsim 1$), the condition to be satisfied is $D_0 \gg R_s^2/R_b u_2$ which is also easily fulfilled for typical values of the parameters.

\begin{figure}
\centering
\includegraphics[width=.45\textwidth]{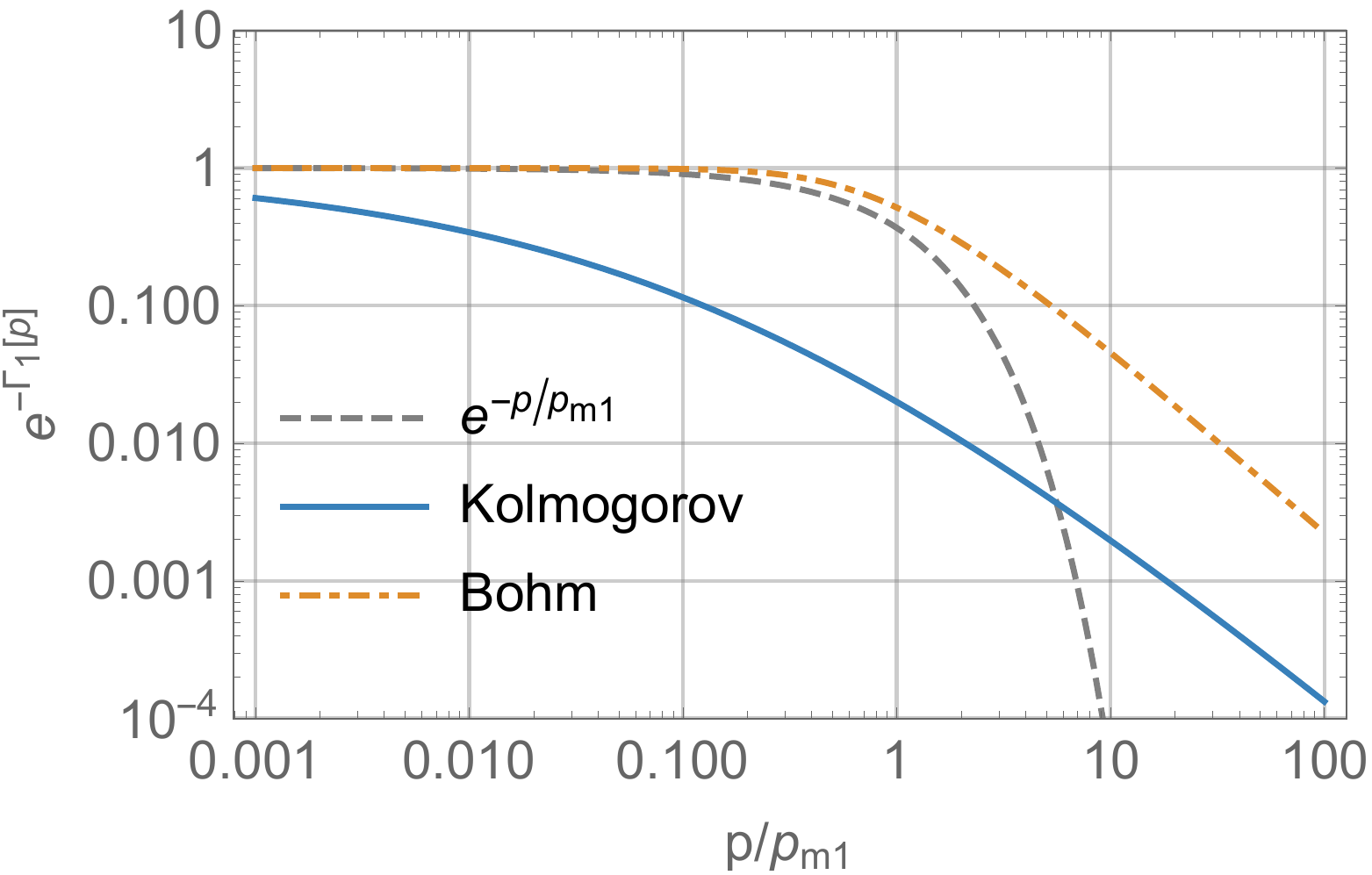}
\includegraphics[width=.45\textwidth]{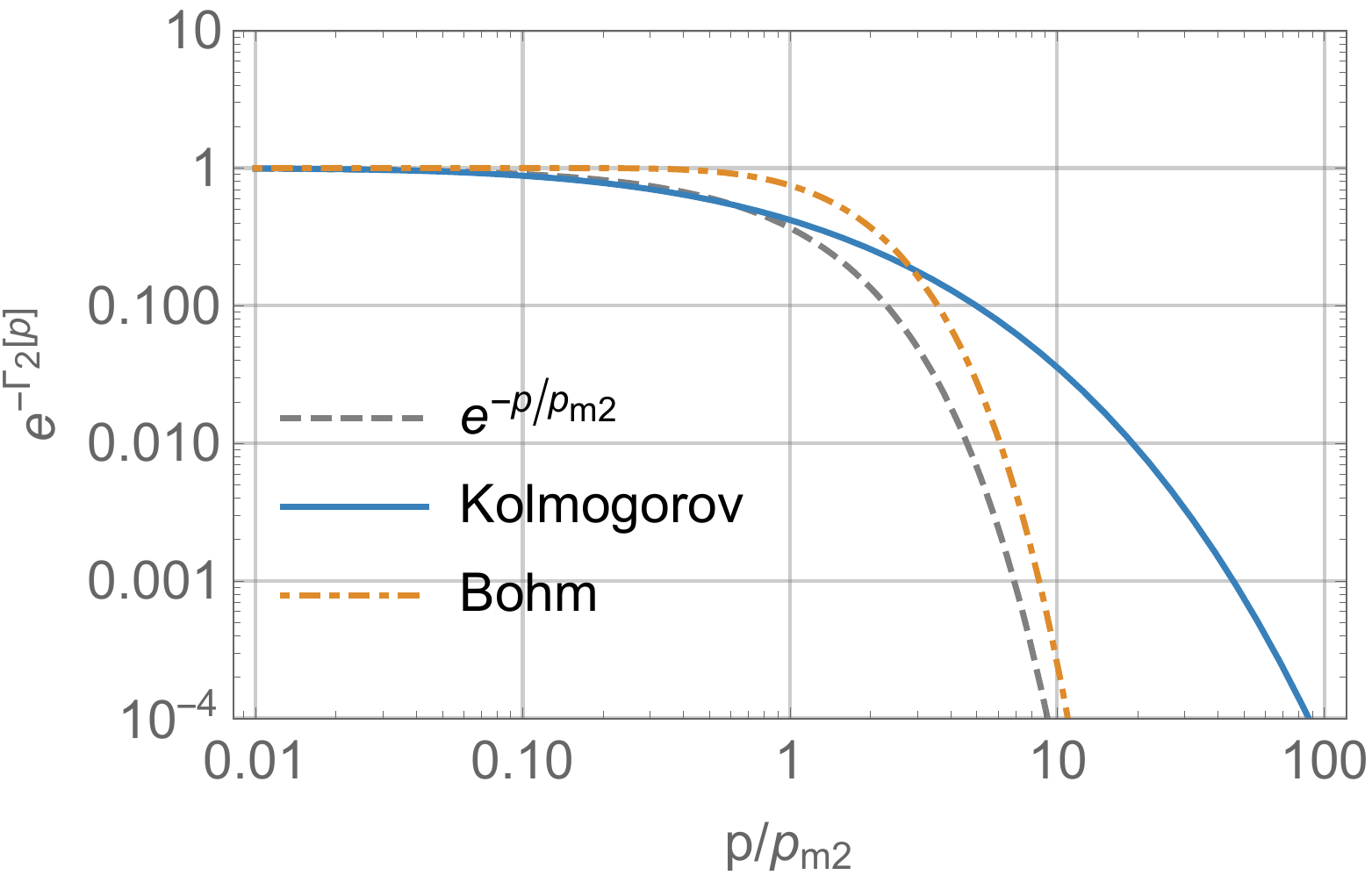}
\caption{Exponential functions $e^{-\Gamma_1(p)}$ (upper panel) and $e^{-\Gamma_2(p)}$ (lower panel) for the case of Kolmogorov and Bohm diffusion. For comparison the simple exponential function is also shown (gray dashed line).}
\label{fig:cutoff}
\end{figure}

The solution at the shock is obtained inserting $\phi_{\rm esc}$ from Eq.\eqref{eq:phi_esc} into Eq.\eqref{eq:fs_diff} which gives:
\beq \label{eq:fs_adiabatic}
  p \frac{\partial f_s}{\partial p} = 
  - \frac{3 u_1}{ u_1 -u_2} \left[ \left( 1- \frac{u_2/u_1}{1-e^{\alpha(R_b)}} + \frac{ G(p)}{u_1 R_s^2 f_s} \right) f_s 
  - \frac{Q_0(p)}{u_1} \right]  \,.
\eeq
The solution of Eq.~\eqref{eq:fs_adiabatic} can be expressed in an implicit form as:
\beq	\label{eq:fs_2}
  f_s(p) = s k \left( \frac{p}{p_{\rm inj}} \right)^{-s}  e^{ -\Gamma_1(p)} e^{ -\Gamma_2(p)} \,, 		
\eeq
where $k=\eta_{\rm inj} n_1/(4\pi p_{\rm inj}^3)$ and $s= 3 u_1/(u_1-u_2)$. The solution, Eq. \eqref{eq:fs_2}, contains three terms. The first one is the usual power law $\propto p^{-s}$ that one finds in the plane parallel shock case while the two exponential terms are 
\bea
  \Gamma_1(p) = s \int_{p_{\rm inj},}^{p} \frac{G(p')}{u_1 R_s^2 f_s(p')} \, \frac{dp'}{p'}  \,, \label{eq:Gamma1} \\
  \Gamma_2(p) = \frac{s}{\sigma} \int_{p_{\rm inj},}^{p} \frac{1}{e^{\alpha_2(p',R_b)} - 1} \, \frac{dp'}{p'}  \, \label{eq:Gamma2}
\eea
and contain the information about the geometry of the system. 
The former, $e^{-\Gamma_1}$, accounts for adiabatic losses/gains in the bubble and contains the whole non-linearity of the solution, depending on both $f$ and $f_s$.
The effect of this term can be understood as follows: first we notice that the function $G(p)$ contains only the contribution $G_1$ from the upstream because $u(r>R_s) \propto r^{-2}$, hence $G_2=0$ (see Eq.~\eqref{eq:G2}) because the  radial expansion of the gas in the shocked wind region is exactly compensated by the velocity decrease. In other words, there are no adiabatic losses. Since the wind velocity is spatially constant in the upstream region, we can write: 
\beq  \label{eq:G1_simply}
  \frac{G_1(\xi,p)}{u_1 R_s^2} =  \frac{2}{3} \int_{0}^{\xi} f_1(\xi,p) \tilde{q}(p)  \xi' d\xi' ,
\eeq
where $\xi=r/R_s$.
Now, if we adopt the solution suitable for a plane parallel shock as a zero order approximation to the real solution, we can estimate the first order correction due to the system geometry. To further simplify the calculation, we consider a spatially constant diffusion coefficient, with a power-law dependence in momentum, $D_1(p)= \kappa_1 p^{\delta_1}$. Hence we assume that $f_s \propto p^{-s}$ and $f_1(\xi,p) = f_s(p) \exp\left[-(1-\xi) \alpha_1 \right]$ where $\alpha_1=u_1 R_s/D_1$. The condition $\alpha_1 = 1$ defines a characteristic momentum 
\beq
  p_{\rm m1} = \left( \frac{u_1 R_s}{\kappa_1} \right)^{1/\delta_1},
\eeq
that characterizes particles able to reach the center of the bubble. Under these simplifying assumptions the function $\Gamma_1$ reduces to
\bea  \label{eq:Gamma1_approx}
  \Gamma_1(p) = s \int_{p_{\rm inj}}^p \frac{dp'}{p'}
    \frac{2}{3}  \int_0^1 \frac{f_1(\xi,p)}{f_s(p)} \tilde{q}(\xi,p) \, \xi d\xi \nn \hspace{2cm}\\
  	= \frac{2 s}{3} \int_{p_{\rm inj}}^p \frac{dp'}{p'} \frac{e^{-\alpha_1}}{\alpha_1^2 } \left\{ s - 3 -\delta_1 (\alpha_1 + 2) + \right. \hspace{1.55cm} \nn \\
    	\left. + e^{\alpha_1} \left[ 3+s (\alpha_1 - 1) + 2\delta_1 - \alpha_1 (3+\delta_1) \right] 
    \right\}	\,. \hspace{0.0cm}
\eea
The shape of $\Gamma_1$ depends mainly on the value of $\delta_1$ and goes to zero for $p \ll p_{\rm m1}$. The function $e^{-\Gamma_1(p)}$ is plotted in Fig.~\ref{fig:cutoff} for the cases of Kolmogorov $(\delta_1=1/3)$ and Bohm $(\delta_1 = 1)$ diffusion. One can see that in both cases a transition occurs at $p=p_{\rm m1}$ but, while in the Bohm case it is very sharp, in the Kolmogorov one it becomes much broader.

The physical meaning of the suppression due to $\Gamma_1$ can also be understood in terms of particle energy gain. For plane parallel shocks, in the test particle limit, the energy gain per cycle is given by $\Delta E/E =4 (u_1-u_2)/(3c)$. In a more general approach $u_1$ and $u_2$ should be replaced by the effective velocities felt by particles upstream, $u_{p1}$, and downstream, $u_{p2}$. In a spherical geometry the effective velocity in the upstream can be written as \citep{Berezhko-Voelk:1997}:
\beq
 u_{p1} = u_1 - \int_{0}^{R_s} dr \frac{\partial(r^2 u)}{\partial r} \frac{f(r,p)}{f_s(p) R_s^2} \,.
 \label{eq:u_p}
\eeq
while $u_{p2} = u_2$ because in the downstream $\partial_r (r^2 u) =0$. 
Using again the approximate expression for the distribution function upstream, $f_1(\xi,p) = f_s(p) \exp\left[-(1-\xi) \alpha_1 \right]$, it is easy to see that for $p\gg p_{\rm m1} \Rightarrow u_p \simeq u_1 \alpha_1/3$, while for for $p\ll p_{\rm m1} \Rightarrow u_p \simeq u_1(1-2/\alpha_1)$. Hence, the energy gain rapidly drops for $p \gg p_{\rm m1}$.

For smaller momenta the asymptotic expression is the same as for a standard shock, but the way that such asymptotic value is approached depends on the spectrum of turbulence, being much more gradual for a Kolmogorov spectrum than for the case of Bohm diffusion.

While $\Gamma_1$ depends on the upstream, $\Gamma_2$ depends only on downstream quantities and produces a cutoff due to particle escape from the bubble boundary. The typical momentum, $p_{\rm m2}$, above which particles can escape the bubble efficiently is defined by the condition $\alpha_2=1$.
In the general case of spatially uniform  diffusion $D_2(p) = \kappa_2 p^{\delta_2}$, such a condition gives
\beq
  p_{\rm m2} = \left[ \frac{u_2 R_s}{\kappa_2} 
  \left(1 - \frac{R_s}{R_b}\right) \right]^{1/\delta_2} \,.
\eeq
The behaviour of $e^{-\Gamma_2}$ is shown in the bottom panel of Figure~\ref{fig:cutoff} for Bohm and Kolmogorov diffusion. Also in this case the Kolmogorov diffusion results in a broader cutoff with respect to the Bohm one but the behaviour for $p<p_{\rm m2}$ is basically identical. The case of Kraichnan diffusion is intermediate between these two.

Summarizing, the maximum energy is limited by two different conditions: the drop of energy gain for $p>p_{\rm m1}$ and the escape from the bubble boundary for $p>p_{\rm m2}$. Since $p_{\rm m1}>p_{\rm m2}$ (at least when $B_2$ simply results from the compression of $B_1$ at the shock), we formally define $p_{\rm m1}\equiv p_{\max}$. 
In addition we have shown that the cutoff depends in a non trivial way by the diffusion properties and its correct shape can only be calculated by solving the full set of equations as we show below.

\subsection{Iterative procedure}
Because the expression for $f_s$ and $f_1$ are implicit, the full solution can be obtained using an iterative procedure. We adopt as a guess function the solution for the plane shock case, namely the one obtained assuming $G=0$ which reads
\begin{flalign} 
  &f_s^{(0)}(p) = s k \left( \frac{p}{p_{\rm inj}} \right)^{-s} e^{-\Gamma_2(p)}  \,, \label{eq:fs_0} \\
  &f_1^{(0)}(\xi,p) = f_s^{(0)}(p) \exp \left[ -\int_{\xi}^1 \frac{u_1 R_s}{D_1} d\xi' \right]		\label{eq:f1_0} \,.
\end{flalign}
Than we compute in sequence $\Gamma_1(p)$, $\tilde{q}(\xi,p)$  and $G_1(\xi,p)$ using Eqs.~\eqref{eq:Gamma1}, \eqref{eq:qtilde} and \eqref{eq:G1_simply}, respectively. At the subsequent steps the iterative expressions are:
\begin{flalign}
  &f_s^{(k+1)} = f_s^{(0)} e^{-\Gamma_1^{(k)}(p)}  \,,    \label{eq:fs_iter} \\
  &f_1^{(k+1)} = f_s^{(k+1)} \exp\left[-\int_{\xi}^1 \frac{u_1 R_s}{D_1} 
  		\left( 1+ \frac{G_1^{(k)}(\xi',p)}{R_s^2 u_1 f_1^{(k)} \xi'^2} \right) d\xi' \right]		\label{eq:f1_iter} \,.
\end{flalign}
The convergence of expressions \eqref{eq:fs_iter} and \eqref{eq:f1_iter} is easily reached within few tens of iterations. When the convergence is reached, we compute $f_2$ using Eq.~\eqref{eq:sol_f2_3}.

\subsection{General properties of the solution}
\label{sec:results}

\begin{figure*}
\centering
{
\includegraphics[width=.45\textwidth]{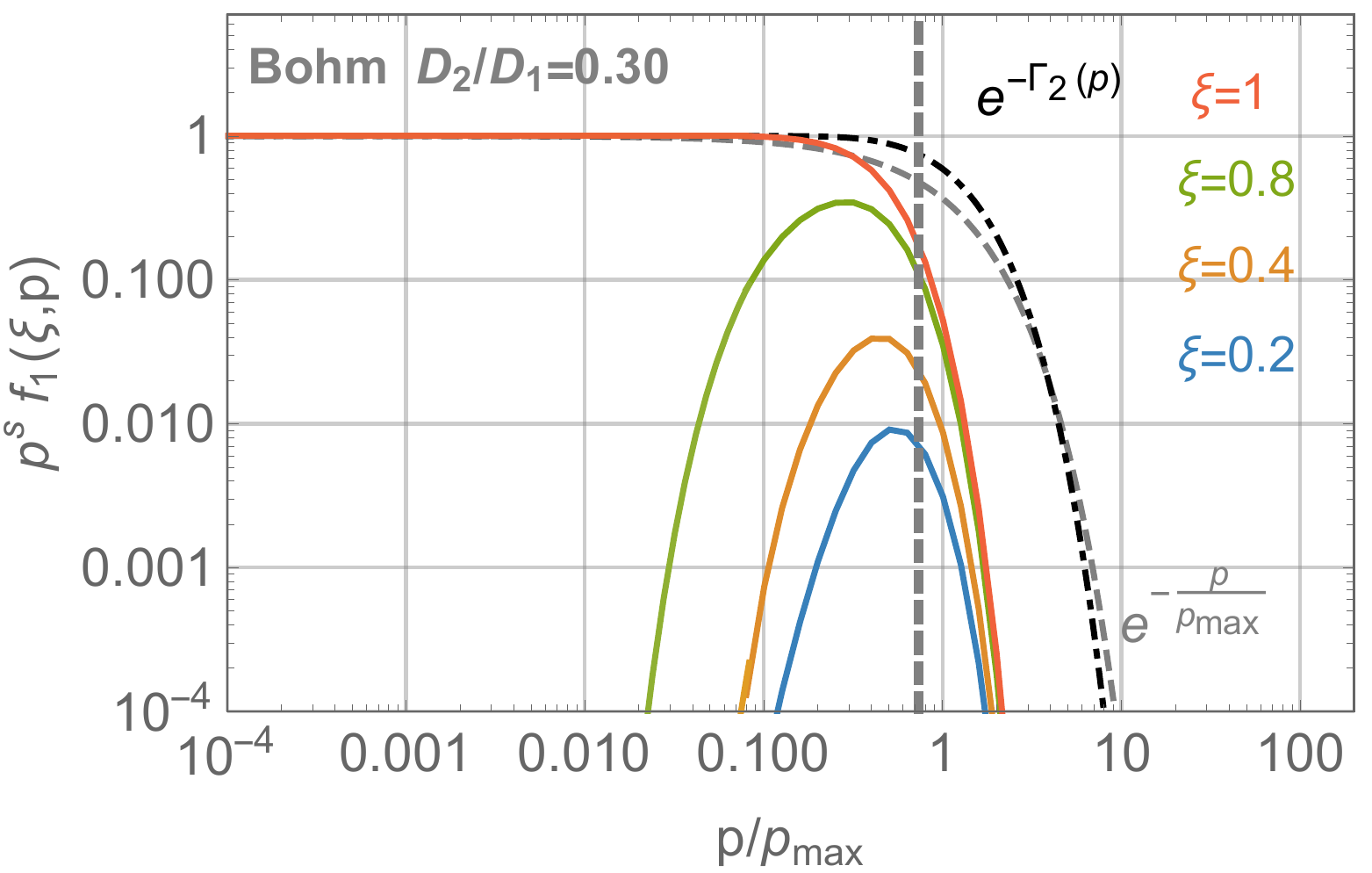}
\includegraphics[width=.45\textwidth]{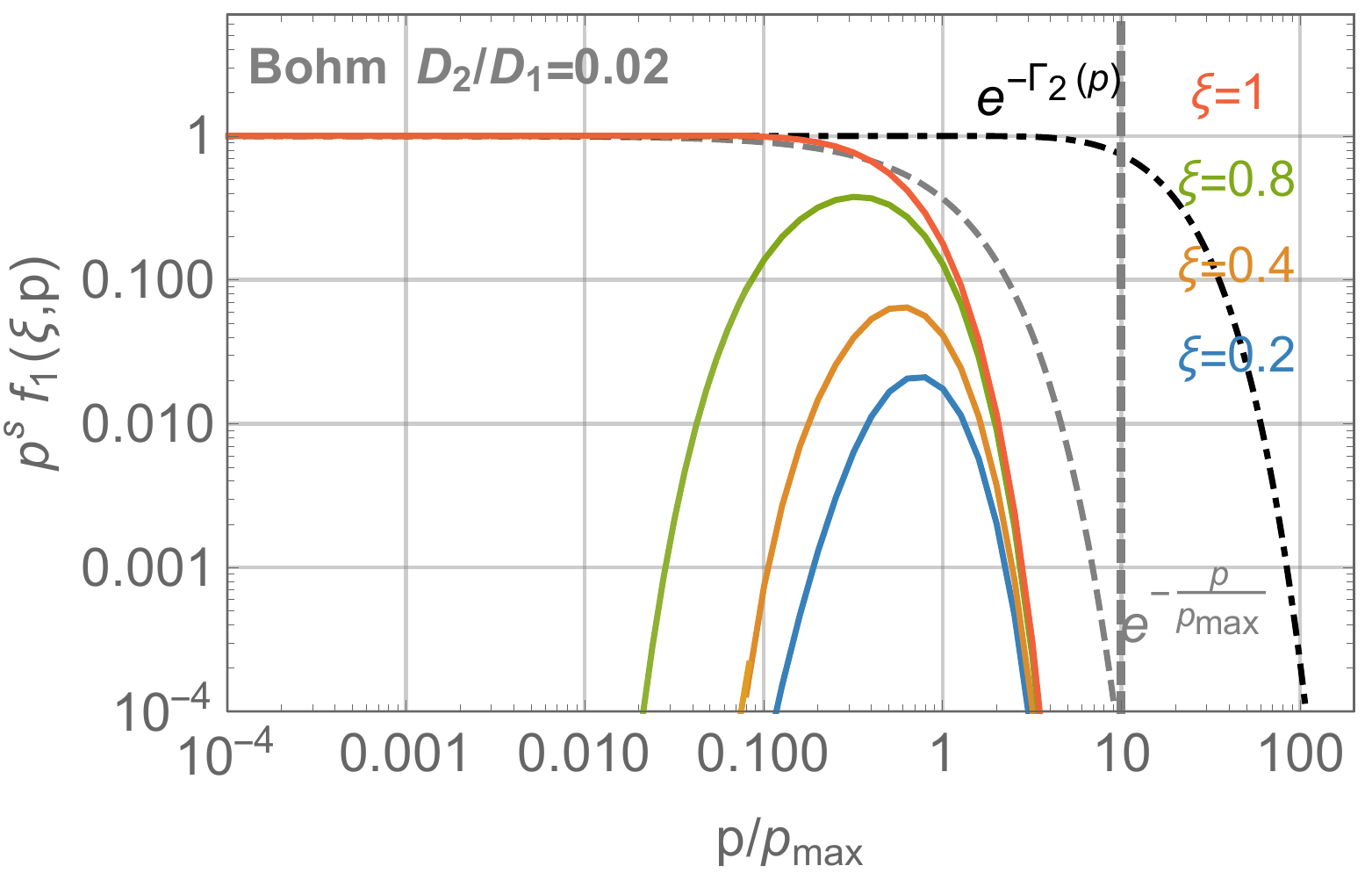}
}
{
\includegraphics[width=.45\textwidth]{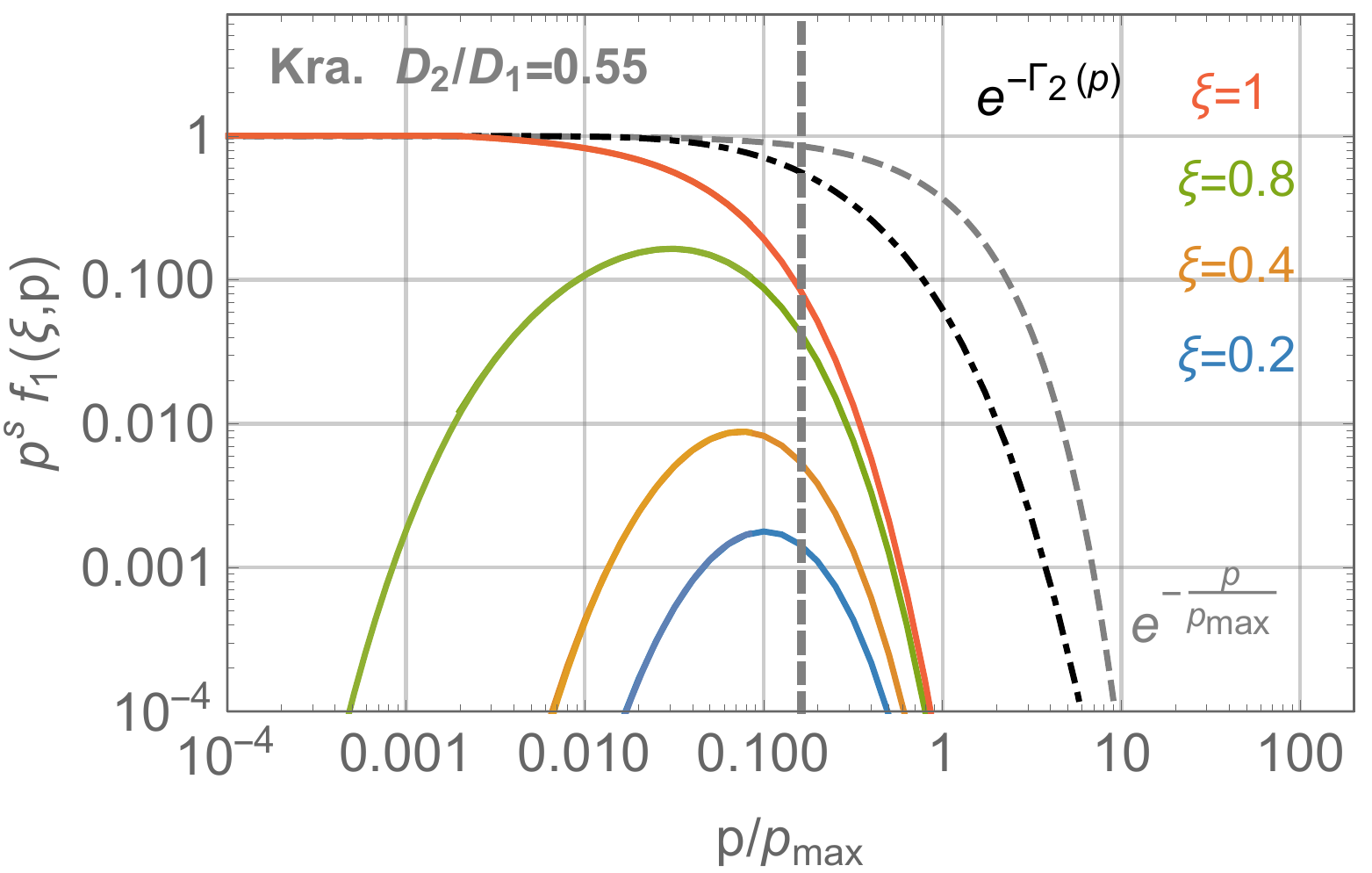}
\includegraphics[width=.45\textwidth]{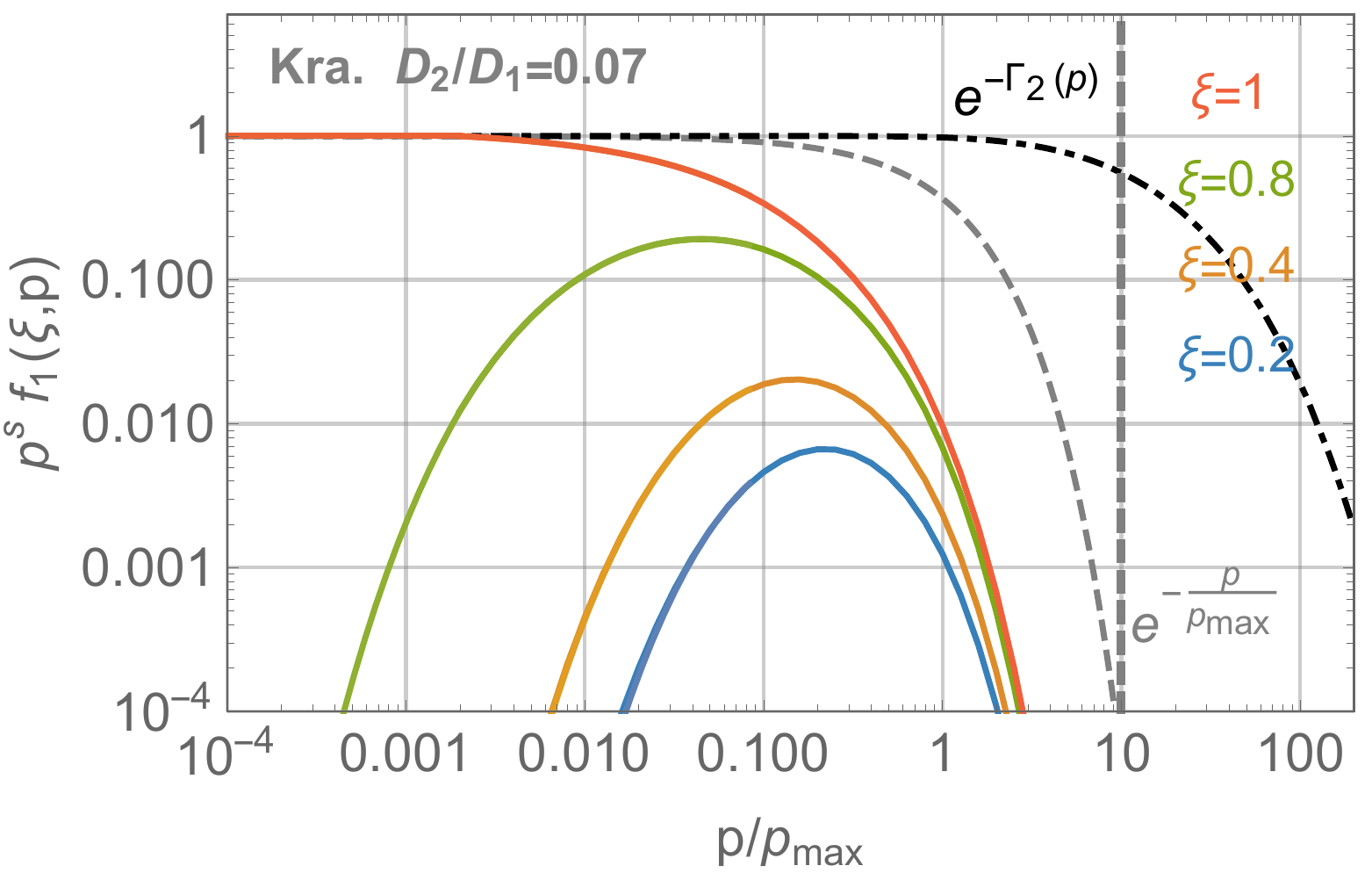}
}
{
\includegraphics[width=.45\textwidth]{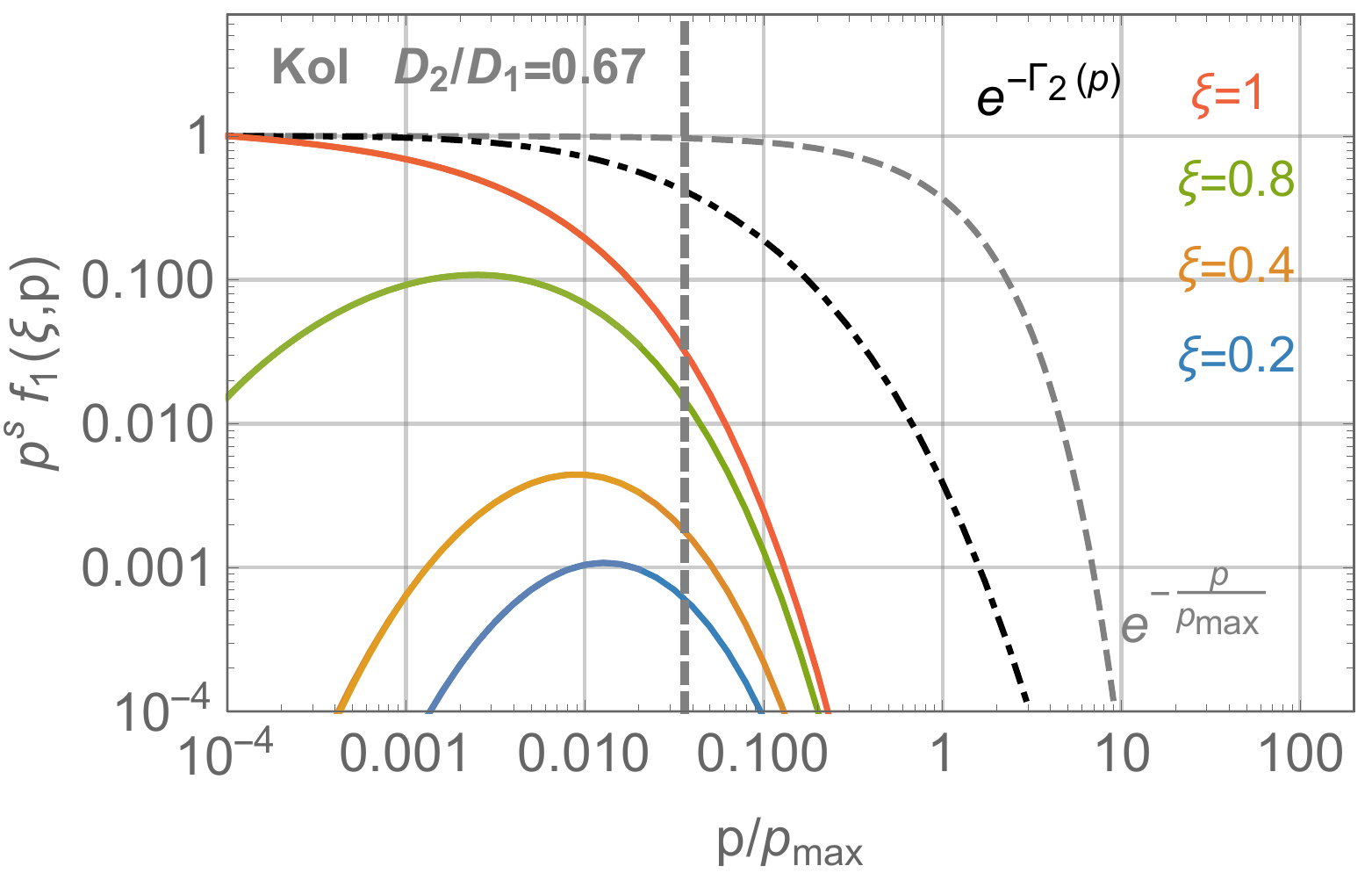}
\includegraphics[width=.45\textwidth]{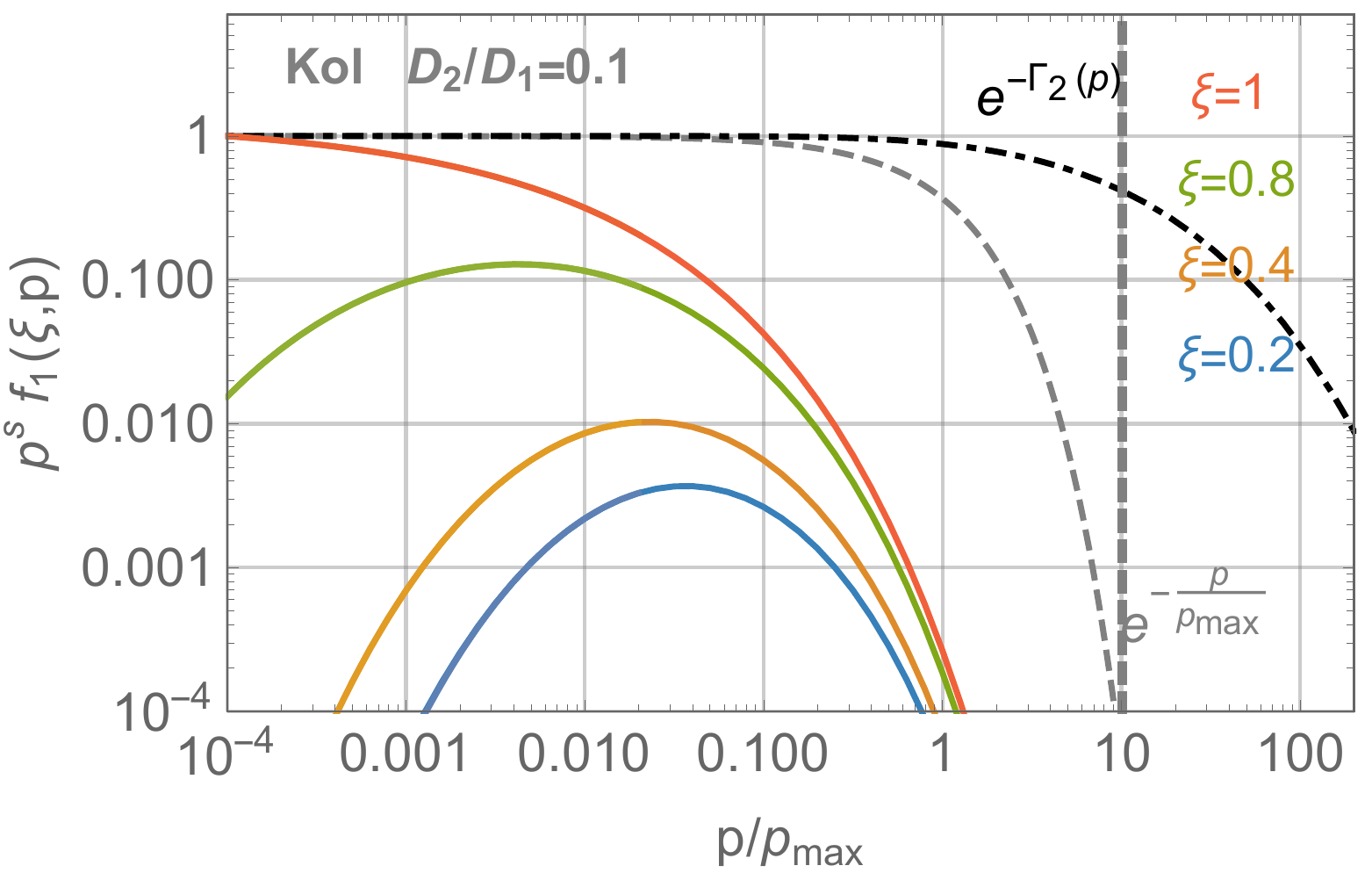}
}
\caption{Particles' spectra multiplied by $p^s$ in the wind region for different distance $\xi$ from the centre ($\xi=1$ is the solution at the termination shock). From top to bottom, the results are shown for different diffusion coefficient: Bohm, Kraichnan and Kolmogorov. Left and right  panels show the solution for different ratios between the diffusion coefficients upstream and downstream. In all plots, the dashed black line shows the zeroth order solution at the shock position when spherical effects are neglected, while the gray dashed line is the simple exponential function shown for comparison. Finally, the vertical dashed line shows the value of $p_{\rm m2}/p_{\rm m1}$.}
\label{fig:f1(p)}
\end{figure*}

Here we illustrate some general considerations concerning the spectrum and spatial distribution of accelerated particles. Figure~\ref{fig:f1(p)} shows the spectrum at different distances upstream of the TS for Bohm, Kraichnan and Kolmogorov diffusion (from top to bottom). The left panels refer to the case when the magnetic field downstream is given only by the compression of the upstream one ($B_2 = \sqrt{11}\,B_1$) as discussed in \S~\ref{sec:MHD_turbulence}, such that $D_2/D_1 = 0.3,~0.55$ and 0.67 for the Bohm, Kraichnan and Kolmogorov cases, respectively. The right panels show, instead,  what happens when the downstream magnetic field is further increased, so as to have $p_{\rm m2} = p_{\rm m1}$.
The curves labelled as $\xi =1$ refer to the location of the TS, while smaller values of $\xi$ refer to the spectrum at locations that are closer to the center of the wind bubble. The dashed black line shows the zeroth order solution at the shock position when the effects due to spherical symmetry are neglected (i.e when $G = 0$) while the grey dashed line show a simple exponential function, reported only for comparison. All spectra are multiplied by $p^{s}$ where $s$ is the slope expected from standard DSA at a planar shock: $s=3\sigma/(\sigma-1)$.
Notice that the momentum is always normalised to the maximum momentum $p_{\max} \equiv p_{\rm m1}$ determined by the upstream conditions and estimated in \S~\ref{sec:diff}. For each case, the corresponding value of $p_{\rm m2}/p_{\rm m1}$ is also reported with a vertical dashed line.

All cases reported in this section are evaluated using typical parameters' values of a massive star cluster, namely: $\dot{M} = 10^{-4} M_{\odot}$, $v_w= 3000$~km s$^{-1}$, $t_b = 10$~Myr,  $n_0 = 1$~cm$^{-3}$ and $\xi_{\rm CR} = \eta_{B} = 0.1$. In addition for the Kraichnan and Kolmogorov cases we fixed the turbulence injection scale at $L_c= 2$~pc.

\begin{figure}
\centering
\includegraphics[width=.45\textwidth]{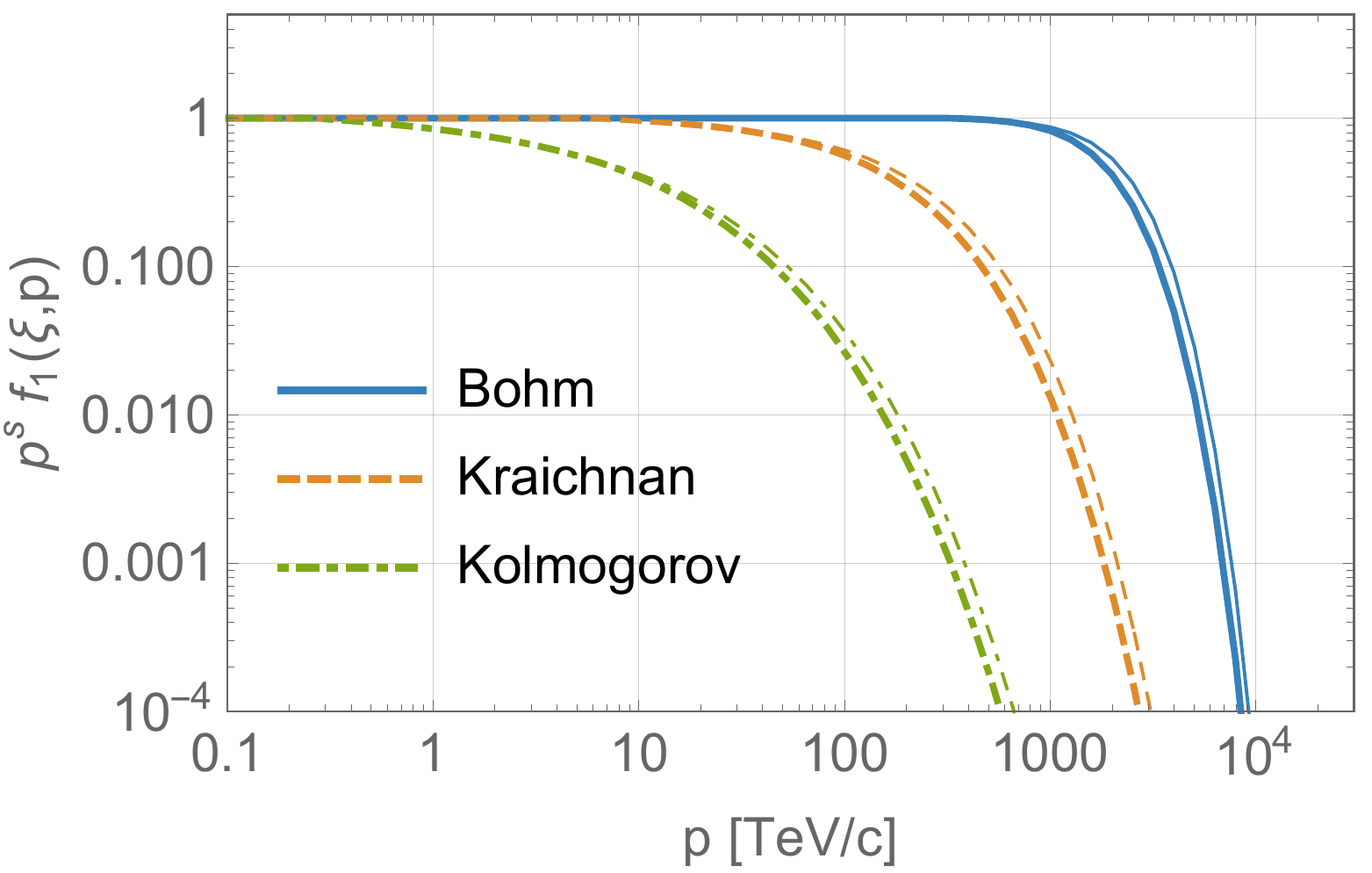}
\caption{Thick lines: distribution function of CR at the shock for different diffusion coefficients. Thin lines: corresponding escaping flux. The results refer to the benchmark case described in the text.}
\label{fig:fs_several_D}
\end{figure}

As expected, particles of lower momenta are spatially concentrated in smaller regions around the TS, while higher energy particles can diffuse farther away from the TS and reach regions of the wind that are close to the star cluster itself. At small $\xi$, basically only particles with $p \sim p_{\max}$ are present. 
The diffusion coefficient determines the shape on the cutoff which is much broader for the Kolmogorov case as compared to the Bohm one. In general the stronger is the energy dependence of $D$, the sharper is the cutoff.
Moreover, the type of diffusion that particles see also determines the effective maximum momentum, defined as the momentum where an appreciable displacement from the power law spectrum at lower energies is visible. The latter approaches $p_{\max}$ only in the case of Bohm diffusion, while the spectrum departs from the power law trend at energies lower than $p_{\max}$ for other types of diffusion. This effect is rather dramatic for Kolmogorov diffusion.
The comparison between different cases can be better appreciated from Figure~\ref{fig:fs_several_D} where the distribution functions at the shock are all plotted together (thick lines).
To provide a quantitative estimate, we define the effective maximum momentum, $\hat{p}_{\max}$, as the momentum where $p^s f_s(p)$ decreases by a factor $1/e$ with respect to a power law extrapolation at lower energies. This quantity is reported in Table~\ref{table:pmax} together with $p_{\rm m1}$ and $p_{\rm m2}$. 
While for the Bohm case $\hat{p}_{\max} \sim 2$~PeV/c, Kraichnan and Kolmogorov diffusion lead to $\hat{p}_{\max} \sim 180$ TeV/c and $\sim 10$ TeV/c, respectively.


Figure~\ref{fig:fs_several_D} also shows the normalised escape flux, $p^s \phi_{\rm esc}(p) /(u_2 R_s^2)$, where $\phi_{\rm esc}(p)$ is given by Eq.~\eqref{eq:phi_esc}. The slightly different shapes of the escape flux and the spectrum of accelerated particles at the TS are limited to the cutoff region, as may be expected in a stationary situation such as the one discussed here.

We stress that, for the reference parameters' values adopted here (corresponding to a wind luminosity $\sim 3\times 10^{38}$ erg s$^{-1}$), the effective maximum energy is in the PeV range for the case of Bohm diffusion and marginally also for Kraichnan diffusion. 
The effective maximum energy could be somewhat increased if additional turbulence is present downstream of the TS, perhaps due to hydrodynamical instabilities. However such an effect is only marginal: in order to illustrate how sensitive $\hat{p}_{\max}$ is to the choice of $D_2$, we arbitrarily decreased the downstream diffusion coefficient (leaving $D_1$ unchanged) so as to have $p_{\rm m2} = 10 p_{\rm m1}$. This happen when $D_2/D_1= 0.02$, 0.05 and 0.1 for Bohm, Kraichnan and Kolmogorov cases, respectively. The corresponding results are shown in the right panels of Figure~\ref{fig:f1(p)}. From these plots, as well as from Table~\ref{table:pmax}, one can see that the impact of decreasing the downstream diffusion coefficient is rather limited: $\hat{p}_{\max}$ increases at most by a factor $\sim 2$ for the  Kolmogorov case, less in the other cases.

\begin{figure}
\centering
\includegraphics[width=.45\textwidth]{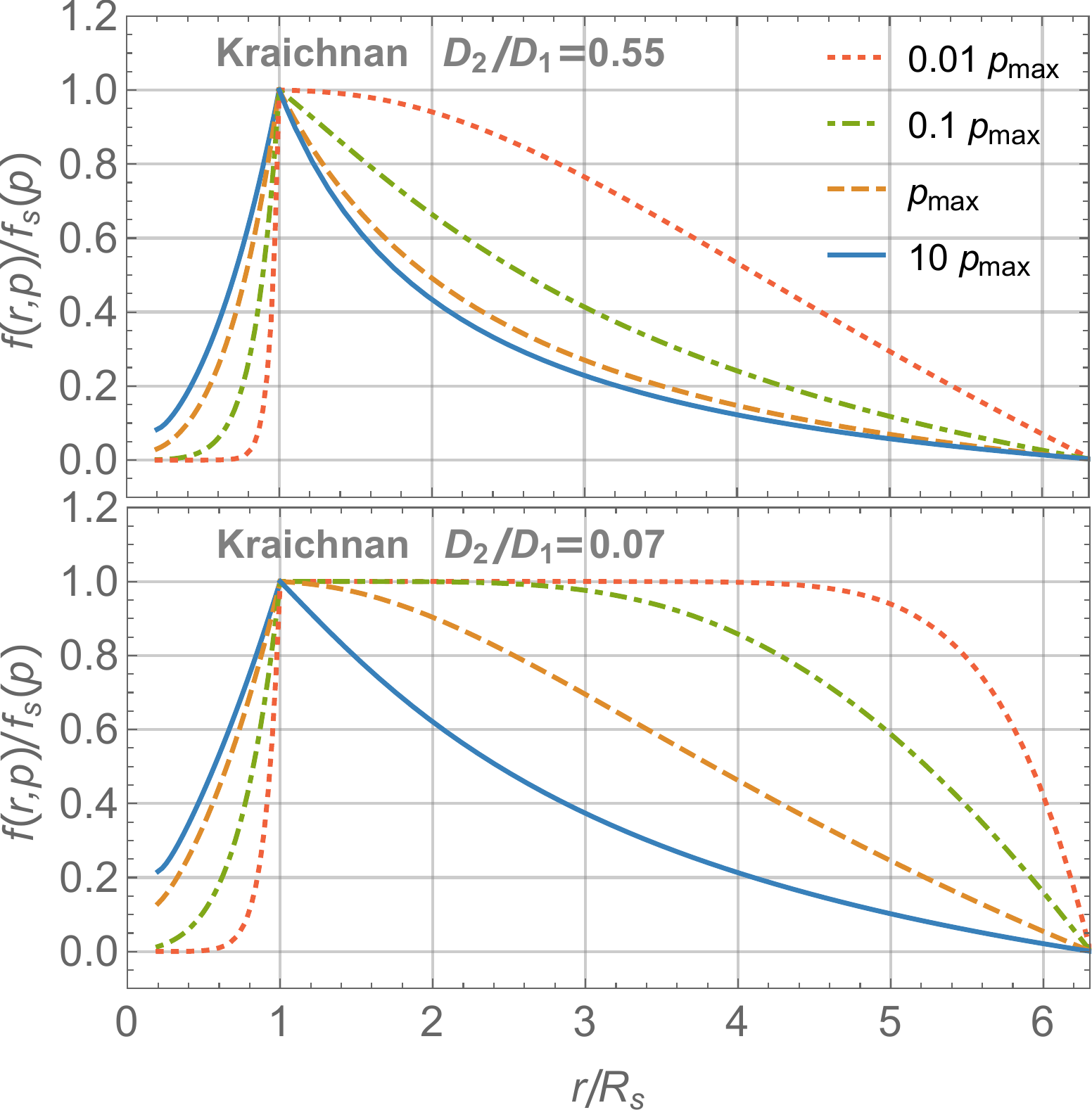}
\caption{Spatial distribution function of CRs normalised at the shock value in the Kraichnan case and for different momenta as shown in the legend. Top and bottom panels show how the results change decreasing the value of $D_2$ from $D_2/D_1= 0.55$ (top) to 0.07 (bottom).}
\label{fig:f(r)}
\end{figure}

The spatial distribution of accelerated particles upstream of the TS is more clearly illustrated in Fig. \ref{fig:f(r)} for the two Kraichnan cases with $D_1/D_2= 0.55$ and 0.07. Larger diffusion coefficient upstream clearly leads high energy particles to diffuse on scales that exceed the radius of the TS, so that they eventually reach the TS on the other side with respect to the central star cluster. When this happens, the effective plasma velocity {\it felt} by particles is $\ll v_w$, hence the energy gain drops to zero and particle acceleration becomes ineffective. The distribution function downstream of the shock becomes flat only for $p \ll p_{\max}$, while for momenta close to $p_{\max}$ the particle density steadily decreases while approaching the bubble boundary.

\begin{table}
\caption{Values of $p_{\rm m1}$, $p_{\rm m2}$ and effective maximum momentum $\hat{p}_{\max}$ for Bohm, Kraichnan and Kolmogorov cases shown in Figure~\ref{fig:f1(p)}.}
\label{table:pmax}
\centering             
\linespread{1.15}\selectfont
\begin{tabular}{c c c c c}
\hline\hline   
Diffusion	&  $D_2/D_1$  &  $p_{\rm m1}$  & $p_{\rm m2}$   &   $\hat{p}_{\max}$	\\    
type	&                      &      [PeV/c]        &  [PeV/c]            &         [PeV/c]              \\
\hline                        
Bohm            & 0.30	& 4.0   & 2.8    &  2.14   \\
 ''             & 0.02	& 4.0   & 40     &  2.80   \\
\hline
Kraichnan       & 0.55	& 3.2	& 0.48   &  0.18   \\
   ''           & 0.07	& 3.2   & 32     &  0.30   \\
\hline
Kolmogorov      & 0.67	& 2.6   & 0.08   &  0.01  \\
   ''           & 0.10	& 2.6   & 26     &  0.02  \\
\hline                                   
\end{tabular}
\linespread{1.0}\selectfont
\end{table}

%

\section{Discussion and conclusions} \label{sec:conc}

There are mainly two reasons for the rising interest of the CR community in star clusters: the first is that if the accelerated particles are extracted from the material expelled by massive stars in the form of stellar winds, the anomalous $^{22}$Ne/$^{20}$Ne abundance ratio, that has been known for quite some time \citep{Binns+:2006}, can be accommodated more easily \citep{Gupta+2020} than by using SNR shocks alone \citep{Prantzos2012}. It should be said that this is all but a trivial conclusion, in that the abundance of $^{22}$Ne in stellar winds depends upon details of the convection of elements in the surface layers of massive stars. But for reasonable models of such phenomenon, it appears that a suitable combination of CRs from massive stars and from SNR shocks should be able to explain observations. 

The second reason for interest in star clusters is that they have been long suspected \cite[]{Cesarsky-Montmerle:1983,webb1985,Gupta+2018,Bykov+2020} to be potential sources of CRs with energies up to the knee. This second aspect turns out to be especially appealing given the many problems encountered by the theory of DSA applied to SNR shock in accounting for such high energies (see for instance \citealt{pierre} and recent reviews \citealt{Blasi2013,Blasi2019}). The possibility of accelerating particles up to the knee in SNRs might be limited to very powerful and rare SN events where the growth of the non resonant instability may be sufficient to reach $\sim 10^{15}$ eV at the beginning of the Sedov-Taylor phase of the shock evolution in the surrounding medium.

In the present article we presented the theory of DSA at the termination shock that arises from the interaction between the collective wind of a star cluster and the surrounding ISM. We solved the stationary transport equation for CRs in spherical symmetry, with a velocity profile that reflects the one expected from the wind region and the bubble region of a star cluster. No restrictions are to be imposed on the spatial and energy dependence of the diffusion coefficient. The solution provides both the spectrum of accelerated particles at any location in the wind and the bubble. The maximum energy arises in a natural way from the transport of particles in the shock region.

As one might expect, the spectrum of accelerated particles, at $p\ll p_{\max}$, is a power law, with the same slope as obtained for a planar shock. This is intuitively clear since the curvature of the shock can affect the particles' diffusion only when the diffusion length is of the same order as the radius of the termination shock. When that happens, the effect of geometry is no longer negligible and one should expect deviations from the standard power law. We showed that the strength of such deviations is very sensitive to the momentum dependence of the diffusion coefficient in the upstream region, being the strongest for weak energy dependence. We investigated in detail three choices, corresponding to a Kolmogorov, Kraichnan and flat spectrum of perturbations. The latter gives rise to Bohm diffusion. In the case of a Kolmogorov spectrum, the deviation from a power law starts a few orders of magnitude in momentum below $p_{\max}$, while the transition is rather sharp at $\sim p_{\max}$ for the case of Bohm diffusion. The Kraichnan case is intermediate between the two but somewhat closer to the Bohm case. In order to quantify the effect of this transition on the particle spectrum we defined an effective maximum momentum, defined as the momentum at which the deviation from a power law extrapolation from lower energies becomes of order $1/e$ on the quantity $p^s f(p)$. These considerations turned out to be highly valuable in terms of assessing the most important point of the article, namely whether PeV energies can be reached at the TS of a star cluster wind.    

The maximum momentum of accelerated particles was found to be in the PeV region for rather bright star clusters, with a wind speed of $\sim 3000$ km/s and mass loss rate of $\sim 10^{-4}~\rm M_\odot yr^{-1}$ corresponding to a kinetic luminosity of $\sim 3 \times 10^{38}$ erg s$^{-1}$. The dependence of $p_{\max}$ on the mass loss rate is almost linear and we notice that in some cases values up to $\sim 10^{-3}~\rm M_\odot yr^{-1}$ have been inferred \citep{Stevens-Hartwell:2003}, leading to maximum energies up to $\sim 10$ times the values reported here. 
The dependence of $p_{\max}$ upon the wind velocity is even stronger, hence for velocities appreciably lower than $\sim 3000$ km s$^{-1}$, $p_{\max}$ rapidly decreases to an extent that depends on the spectrum of perturbations. Although $p_{\max}$ can be in the PeV range, as discussed above, the effective maximum momentum can be much lower as a result of spherical symmetry, especially for Kolmogorov diffusion. This issue is less pressing for Kraichnan and Bohm diffusion, where the spectrum shows an effective suppression at $\sim 0.1-1 p_{\max}$. Hence, having a correct understanding of the magnetic turbulence in this environment is of the utmost importance. We also stress that at the large CR energy we are interested in (resonant scales comparable with $L_c$) the effects of anisotropic development of turbulence \cite[]{GS1,GS2} should not be overwhelmingly important.

We also showed that for a reasonable choice of parameters, the maximum momentum is determined by the conditions upstream of the termination shock, namely in the cold wind. This might appear counter intuitive, since no escape is possible from the upstream region, due to geometry. However, it can be easily understood in terms of the effective plasma velocity that the particles experience in the wind. Such velocity is close to $v_w$ when the diffusion length is much smaller than $R_{s}$ (low momenta). However it decreases for higher momenta, and eventually becomes close to zero when the diffusion length exceeds $R_{s}$. This can also be seen in terms of energy gain per cycle, which decreases when the diffusion length of the particles becomes comparable to $R_{s}$. 

In principle somewhat larger values of $p_{\rm max}$ can be obtained if additional turbulence exists downstream of the termination shock, for instance excited through some kind of hydrodynamic instability. However, as one could expect, this reflects in only mild increases in the maximum momentum, since the latter is more strongly constrained by the upstream conditions.  
In conclusion, star clusters are potential sources of CR protons at the knee only for very bright and relatively uncommon objects. Even for the star clusters for which the maximum energy is in the PeV region, the shape of the spectrum close to $p_{\max}$ is such that it may result in an early suppression, for the case of Kolmogorov-like diffusion. The question of whether these objects can contribute an appreciable flux of light CRs in the knee region should then be addressed using observations of X-rays and very high energy gamma rays, which can provide valuable information on the conditions at the TS.

\section*{Acknowledgements}
The research activity of PB and GM was partially funded through support Grant ASI/INAF n. 2017-14-H.O; GM also was funded through Grants SKA-CTA-INAF 2016 and INAF-Mainstream 2018. The research activity of EP was supported by Villum Fonden under project n. 18994.

\section*{Data Availability}
There are no new data associated with this article.
 




\bibliographystyle{mnras}
\bibliography{biblio} 






\bsp	
\label{lastpage}
\end{document}